\documentclass[twocolumn]{aastex62}

\hypersetup{linkcolor=blue,citecolor=blue,filecolor=blue,urlcolor=blue}

\usepackage{array, makecell}
\hypersetup{colorlinks, citecolor=blue, urlcolor=blue, linkcolor=blue}
\usepackage{rotating}
\usepackage{array}
\usepackage{amsmath,amsfonts,amsthm}
\usepackage{multirow}
\usepackage{indentfirst}

\graphicspath{{./}{figures/}}

\received{}
\revised{}
\accepted{}
\submitjournal{ApJ}

\shorttitle{What determines active region coronal plasma composition?}
\shortauthors{Mihailescu et al.}

\begin{document}

\title{What determines active region coronal plasma composition?}

\correspondingauthor{Teodora Mihailescu}
\email{teodora.mihailescu.19@ucl.ac.uk}

\author[0000-0001-8055-0472]{Teodora Mihailescu}
\affil{Mullard Space Science Laboratory, University College London, UK}

\author[0000-0002-0665-2355]{Deborah Baker}
\affil{Mullard Space Science Laboratory, University College London, UK}

\author[0000-0002-0053-4876]{Lucie M. Green}
\affil{Mullard Space Science Laboratory, University College London, UK}

\author[0000-0002-2943-5978]{Lidia van Driel-Gesztelyi}
\affil{Mullard Space Science Laboratory, University College London, UK}
\affil{Observatoire de Paris, LESIA, France}
\affil{Konkoly Observatory, Hungarian Academy of Sciences, Hungary}

\author[0000-0003-3137-0277]{David M. Long}
\affil{Mullard Space Science Laboratory, University College London, UK}

\author[0000-0002-2189-9313]{David H. Brooks}
\affil{College of Science, George Mason University, VA, USA}

\author[0000-0003-0774-9084]{Andy S. H. To}
\affil{Mullard Space Science Laboratory, University College London, UK}




\begin{abstract}
The chemical composition of the solar corona is different from that of the solar photosphere, with the strongest variation being observed in active regions (ARs). Using data from the Extreme Ultraviolet (EUV) Imaging Spectrometer (EIS) on Hinode, we present a survey of coronal elemental composition as expressed in the first ionisation potential (FIP) bias in 28 ARs of different ages and magnetic flux content, which are at different stages in their evolution. We find no correlation between the FIP bias of an AR and its total unsigned magnetic flux or age. However, there is a weak dependence of FIP bias on the evolutionary stage, decreasing from 1.9-2.2 in ARs with spots to 1.5-1.6 in ARs that are at more advanced stages of the decay phase. FIP bias shows an increasing trend with average magnetic flux density up to 200 G but this trend does not continue at higher values. The FIP bias distribution within ARs has a spread between 0.4 and 1. The largest spread is observed in very dispersed ARs. We attribute this to a range of physical processes taking place in these ARs including processes associated with filament channel formation. These findings indicate that, while some general trends can be observed, the processes influencing the composition of an AR are complex and specific to its evolution, magnetic configuration or environment. The spread of FIP bias values in ARs shows a broad match with that previously observed in situ in the slow solar wind.
\end{abstract}



\section{Introduction}
The composition of the Sun's plasma is a key indicator of important physical processes at play in the solar atmosphere, such as heating or mass and energy transport. While the photospheric plasma composition is relatively well determined and constant across the Sun's surface and in time \citep{Asplund2009}, the coronal plasma composition is variable and can be different from the photospheric values \citep{Meyer1985}. The presence of this variability is also supported by in-situ measurements of the chemical composition of the solar wind, which also show a variable composition \citep{vonSteiger2000}.

The abundance variation of an element is strongly dependent on its first ionisation potential \citep[FIP; ][]{Meyer1985, Meyer1985a} and not on other parameters such as mass or charge \citep{Meyer1991}. Elements with a low FIP such as Si, Fe, Mg and Ca are enhanced in the corona, compared to high-FIP elements such as S, Ar, Ne and O, which maintain their photospheric abundances. This is called the FIP effect. To characterise the degree of enhancement of low-FIP elements in the corona and how it changes with time, we use the FIP bias parameter:

\begin{equation}\label{FIPformula}
    \text{FIP}_\text{bias}=\frac{\text{coronal elemental abundance}}{\text{photospheric elemental abundance}}.
\end{equation}

So far, the theoretical model that best explains the FIP effect is the ponderomotive force model - initially proposed by \citet{Laming2004} with further developments being detailed in the review by \citet{Laming2015}. Ponderomotive forces arise from the effects of wave refraction in an inhomogenous plasma. The model proposes that, in the solar atmosphere, Alfv\'en waves can be generated in the corona, travel to lower altitudes and refract in the high density gradient of the chromosphere. This change in wave direction can exert a ponderomotive force on the ions in the plasma, acting as an agent to separate them from the neutrals. Ions then travel in the direction of high wave energy density, so a higher wave energy density in the corona leads to a stronger FIP effect. This effect is amplified in closed loops where nanoflares can give rise to coronal Alfv\'en waves. These Alfv\'en waves gradually refract at the loop chromospheric footpoints, until they undergo total internal reflection and travel back into the corona. The model proposes that coronal Alfv\'en waves are naturally at resonance with the loop so they travel repeatedly between footpoints, continuously driving the fractionation. Typical timescales of this process are on the order of hours to a couple of days.  

The strongest FIP effect is observed in active regions (ARs). The plasma composition in an AR varies throughout its evolution and is modulated by different processes during the emergence and decay phases \citep{Widing2001a, baker2015a, Ko2016, Baker2018}. A study of 4 emerging ARs at solar minimum found the emerging flux initially exhibited plasma with photospheric abundances \citep{Widing2001a}. The observations, based on Skylab spectroheliograms, used the Mg/Ne ratio ($\log{\mathrm{T}}\approx 5.6-5.7$, i.e. $T \approx 300,000-500,000$ K) as a measure of the FIP bias. The results suggest that emerging AR loops bring up material from the photosphere into the corona. As flux emergence continued, the FIP bias gradually increased almost linearly with AR age for 3-4 days in the analysed large ARs.

A more recent study, however, showed that this FIP bias increase does not continue in the later stages as suggested by initial Skylab observations \citep{Widing2001a}, but rather it starts to decrease once the AR goes through its middle and late decay phases \citep{baker2015a}. The observations, from the Hinode EUV Imaging Spectrometer \citep[EIS; ][]{Culhane2007a}, used the Si {\footnotesize X} 258.38 \AA{}/S {\footnotesize X} 264.23 \AA{} ($\log{\mathrm{T}}\approx 6.2$, i.e. $T \approx 1.5$ MK) line ratio as a FIP bias measure \citep[see][for a complete description of the method]{Brooks2011, Brooks2015}. In the studied AR, the decay phase was dominated by a global decrease in FIP bias. Small bipoles emerged within and around the boundary of supergranular cells, as the magnetic field got progressively more dispersed. The small newly emerged loops contain photospheric plasma and their reconnection with older AR loops brings this photospheric material upwards into the corona, leading to plasma mixing. The mixing timescales are shorter than the fractionation timescales so the overall FIP bias decreases. Another study (using a similar FIP bias diagnostic) followed the decay phase of another large AR and found that FIP bias values decrease in the decay phase and eventually settle around the FIP bias value of 1.5, corresponding to the FIP bias value of the surrounding quiet Sun \citep{Ko2016}. 

A subsequent study, following the temporal evolution of coronal plasma composition within seven emerging flux regions inside a coronal hole (CH), found that FIP bias increases in the emergence and early decay phases, before decreasing in the middle and late decay phase \citep{Baker2018}. \citet{Baker2018} proposed that the FIP bias increase in the emergence phase is driven by the fractionation process \citep{Laming2004, Laming2015} and transport of fractionated plasma into the corona, while the FIP bias decrease in the late decay phase is linked to the composition of the surrounding corona and the rate of reconnection with this surrounding magnetic field.

As well as temporal variation, ARs show spatial variation in FIP bias. The highest FIP bias values are observed at AR loop footpoints \citep{Baker2013}, indicating that this is where the fractionation process takes place \citep[as proposed by][]{Laming2004}. Traces of high FIP bias are observed along some of the AR loops, indicating plasma starting to mix along loops \citep{Baker2013}. Flux cancellation along the AR main polarity inversion line (PIL), and the associated flux rope formation leads to lower FIP bias levels \citep{Baker2013, Baker2022}. Lower FIP bias levels were also found in the part of an AR where two failed eruptions occured \citep{baker2015a}. The coronal plasma above the cool umbra of a very large sunspot was found to have photospheric composition, while the coronal loops rooted in the penumbra showed fractionated plasma, with the highest FIP bias values (3-4) being observed in the loops that connect within the AR \citep{Baker2021}.

These previous studies found trends in how FIP bias evolves and is distributed within ARs, and it would be interesting to analyse whether the observed trends hold for all ARs. Key questions include: are the composition trends similar for ARs of different sizes? Do larger ARs reach higher FIP bias values? How does the FIP bias of an AR evolve in the very late stages of the decay phase?

In this survey, we analyse plasma composition data from 28 ARs from 3 full Sun EIS scans to explore how FIP bias relates to their evolutionary stage and magnetic configuration. The first full Sun FIP bias map was initially used to investigate potential slow solar wind sources \citep{Brooks2015}, as well as compare in situ solar wind composition data to its source region composition measured by EIS \citep{stansby2020c}. The 28-AR dataset contains ARs of all evolutionary stages, from emergence to decay, including very dispersed ARs that are in a more developed stage than the ones analysed in previous studies. We look at individual case studies to investigate how particular aspects of an AR (e.g. filament/filament channel formation, flux cancellation) can influence coronal plasma composition. 

\begin{figure*}
    \centering
    \makebox[\textwidth][c]{\includegraphics[width=1.0\textwidth]{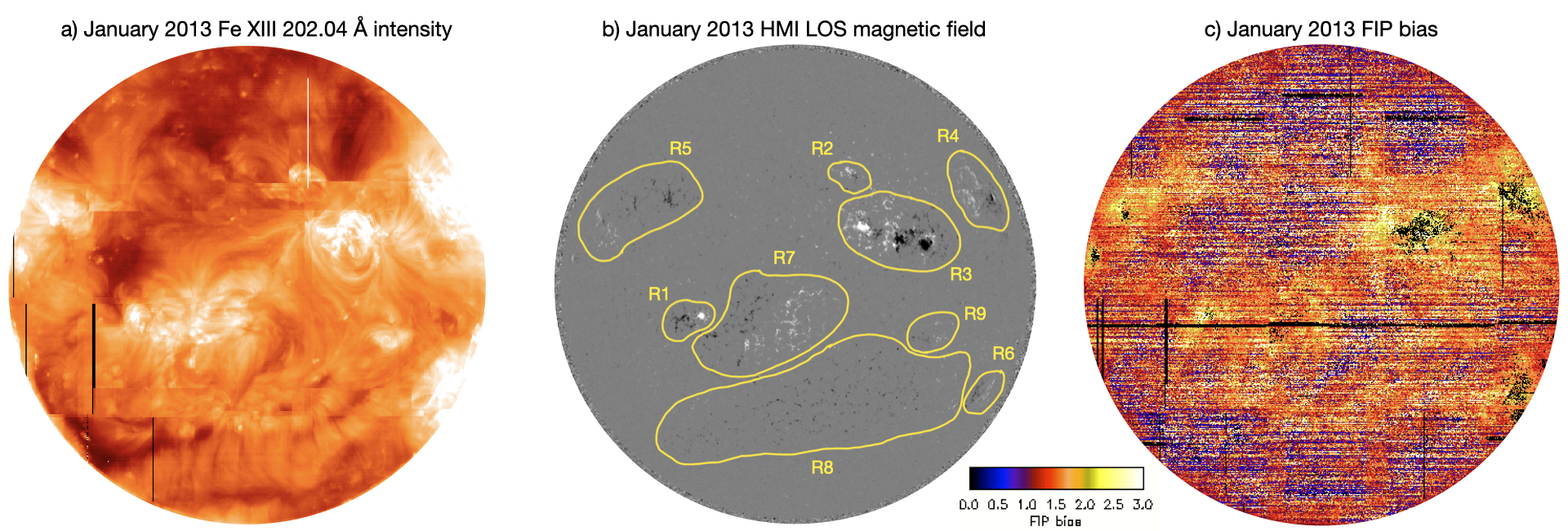}}
    \makebox[\textwidth][c]{\includegraphics[width=1.0\textwidth]{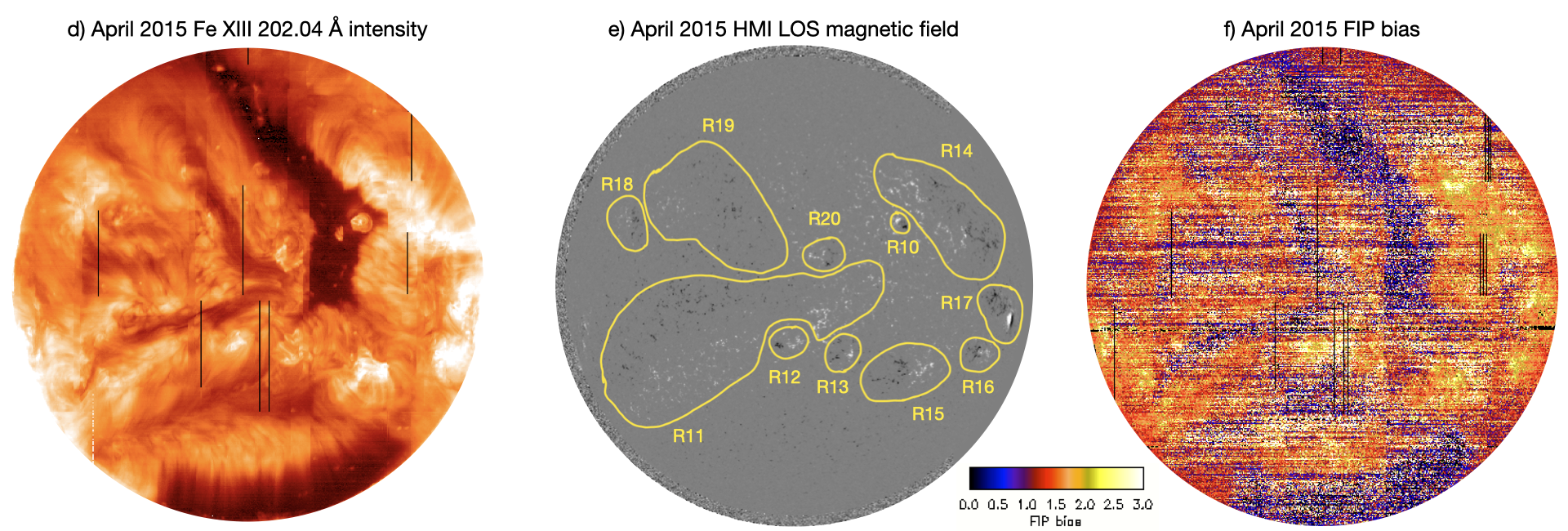}}
    \makebox[\textwidth][c]{\includegraphics[width=1.0\textwidth]{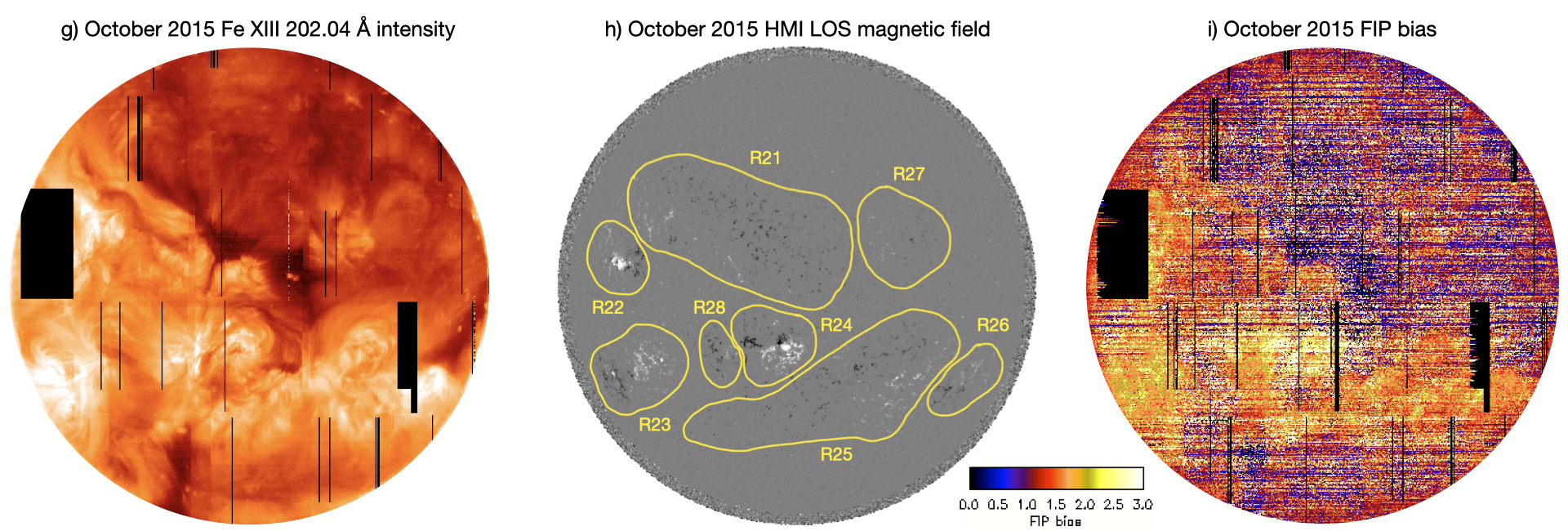}}
    \caption{Full Sun maps contructed from: a, d, g) Hinode/EIS Fe {\footnotesize XIII} 202.04 \AA{} intensity; b, e, h) HMI line of sight photospheric magnetic field strength with overlaid active region contours (in yellow); c, f, i) Hinode/EIS FIP bias. The black boxes present in the FIP bias and Fe {\footnotesize XIII} 202.04 \AA{} intensity maps represent gaps in the EIS data. Individual active region contours overlaid on Hinode/EIS Fe {\footnotesize XIII} 202.04 \AA{} intensity, HMI line of sight photospheric magnetic field strength and Hinode/EIS FIP bias maps are shown in Appendix \ref{IndividualContours}.}
    \label{FDscans}
\end{figure*}

\begin{deluxetable*}{ccccccccccc}
\tabletypesize{\footnotesize}
\centering
\tablecolumns{2}
\tablehead{\colhead{Region} & \colhead{NOAA} & \colhead{Type} & \colhead{Age} & \colhead{Evolutionary} & \colhead{Total magnetic} & \colhead{Mean magnetic} & \colhead{Median} & \colhead{Skewness} & \colhead{Spread} & \colhead{Percentage}\\
\colhead{} & \colhead{number} & \colhead{} & \colhead{(days)} & \colhead{stage} & \colhead{flux ($\times 10^{20}$ Mx)} & \colhead{flux density (G)} & \colhead{FIP bias} & \colhead{coefficient} & \colhead{} & \colhead{FIP bias $> 4$}}
\tablecaption{General characteristics of active regions presented in Figure \ref{FDscans}: active region R code (1-28), NOAA active region number (if available), type (activity nest or simple bipolar region), active region age (and the age of the activity nest, if the active region is part of a nest), evolutionary stage, total magnetic flux content at the time of the scan, average magnetic flux density, median FIP bias (50th percentile), Kelly's skewness coefficient, spread (defined as the difference between the 75th and the 25th percentiles of the FIP bias distribution) and percentage of pixels with a FIP bias value higher than 4 for the active regions in the study. \label{ARcharacteristics}}
\startdata
R1 & 11658 & Simple & 5 & Spots & $59 \pm 0.07$ & $200 \pm 0.2$ & 1.9 & 0.12 & 0.6 & 0.1\\
R2 & 11656 & Simple & 11 & Decayed & $15 \pm 0.04$ & $145 \pm 0.4$ & 1.7 & 0.15 & 0.5 & 0.6\\
R3 & 11654 & Nest & 11 (63) & Spots & $364 \pm 0.17$ & $222 \pm 0.1$ & 1.9 & 0.21 & 0.6 & 0.9\\
R4 & 11652 & Simple & 13-26 & Spots & $59 \pm 0.10$ & $103 \pm 0.2$ & 1.9 & 0.15 & 0.5 & 0.0\\
R5 & N/A & Simple & 39 & Dispersed & $52 \pm 0.09$ & $100 \pm 0.2$ & 1.7 & 0.19 & 0.6 & 0.7\\
R6 &  11650 & Simple & 13-26 & Decayed & $16 \pm 0.06$ & $83 \pm 0.3$ & 1.7 & 0.15 & 0.4 & 0.2\\
R7 & N/A & Simple & 53 & Dispersed & $123 \pm 0.13$ & $120 \pm 0.1$ & 1.6 & 0.18 & 0.6 & 0.5 \\
R8 & N/A& Nest & 189 (244) & Filament channel & $70 \pm 0.12$ & $81 \pm 0.1$ & 1.5 & 0.25 & 0.8 & 1.0\\
R9 & 11657 & Simple & 13-24 & Decayed & $10 \pm 0.04$ & $87 \pm 0.4$ & 1.6 & 0.18 & 0.6 & 0.1\\
\hline
R10 & 12317 & Simple & 0.5 & Emerging & $19 \pm 0.04$ & $233 \pm 0.5$ & 1.7 & 0.16 & 0.7 & 0.7\\
R11 & N/A & Nest & 120 (214) & Filament channel &$197 \pm 0.18$ & $95 \pm 0.1$ & 1.6 & 0.25 & 0.8 & 2.0\\
R12 & 12316 & Simple & 6-21 & Decayed & $20 \pm 0.05$ & $147 \pm 0.4$ & 2.0 & 0.27 & 0.8 & 3.0\\
R13 & 12314 & Simple & 8-20 & Decayed &$14 \pm 0.04$ & $141 \pm 0.4$ & 1.7 & 0.19 & 0.6 & 0.5\\
R14 & 12310 & Nest & 8-25 (86) & Dispersed & $95 \pm 0.12$ & $106 \pm 0.1$ & 1.8 & 0.16 & 0.5 & 0.3\\
R15 & N/A & Simple & 24 & Decayed & $41 \pm 0.08$ & $119 \pm 0.2$ & 1.8 & 0.19 & 0.7 & 2.0\\
R16 & 12315 & Simple & 11-19 & Decayed & $18 \pm 0.05$ & $113 \pm 0.3$ & 1.8 & 0.10 & 0.5 & 0.1\\
R17 & 12305 & Simple & 12-19 & Spots & $64 \pm 0.05$ & $169 \pm 0.2$ & 1.9 & 0.21 & 0.5 & 0.5\\
R18 & N/A & Simple & 3-17  & Decayed & $13 \pm 0.04$ & $99 \pm 0.4$ & 1.7 & 0.13 & 0.5 & 0.4\\
R19 & N/A & Nest & 58-76 (111) & Filament channel & $39 \pm 0.09$ & $84 \pm 0.2$ & 1.5 & 0.23 & 0.8 & 2.0\\
R20 & N/A& Simple & 7-20 & Decayed &$15 \pm 0.05$ & $125 \pm 0.4$ & 1.6 & 0.29 & 0.8 & 2.0\\
\hline
R21 & N/A& Nest & 29 (183) & Filament channel & $143 \pm 0.16$ & $97 \pm 0.1$ & 1.5 & 0.25 & 1.0 & 4.0\\
R22 & 12436& Simple & 30-44 & Spots & $100 \pm 0.10$ & $166 \pm 0.2$ & 1.9 & 0.08 & 0.6 & 0.4\\
R23 & N/A & Simple & 27 & Spots & $107 \pm 0.10$ & $115 \pm 0.1$ & 1.9 & 0.10 & 0.4 & 1.0\\
R24 & 12434 & Simple & 27 & Spots & $188 \pm 0.12$ & $213 \pm 0.1$ & 2.2 & 0.23 & 1.0 & 4.0\\
R25 & N/A & Simple & 63-79 & Dispersed & $107 \pm 0.13$ & $104 \pm 0.1$ & 1.7 & 0.23 & 0.7 & 1.0\\
R26 & N/A & Simple & 38-51 & Dispersed & $28 \pm 0.07$ & $97 \pm 0.2$ & 1.7 & 0.09 & 0.4 & 0.1\\
R27 & N/A & Simple & 35 & Dispersed & $13 \pm 0.05$ & $79 \pm 0.3$ & 1.4 & 0.34 & 0.9 & 2.0\\
R28 & N/A & Simple & 21-25 & Decayed & $21 \pm 0.06$ & $119 \pm 0.3$ & 1.9 & 0.24 & 0.6 & 1.0\\
\enddata
\end{deluxetable*}

\section{Observations}
The dataset used for this study comprises three full Sun EIS scans taken on 16-18 January 2013, 1-3 April 2015 and 18-20 October 2015. In total, these scans cover 28 ARs and provide composition measurements at the time of the scan. The ARs are shown in Figure \ref{FDscans} and their general characteristics are given in Table \ref{ARcharacteristics}.

\subsection{Coronal EUV and Magnetic field observations}
\label{Coronal EUV and Magnetic field observations}
The history of each AR was explored using line of sight magnetogram images from the Solar Dynamics Observatory \citep[SDO; ][]{Pesnell2012} Helioseismic and Magnetic Imager \citep[HMI; ][]{Scherrer2012, Schou2012}, to determine its approximate age and complexity at the time of each EIS scan. EUV images from the Atmospheric Imaging Assembly \citep[AIA; ][]{Lemen2012} instrument, particularly  in the 171 \AA{} ($\log{\mathrm{T}} \approx 5.8$; quiet corona and upper transition region) and 193 \AA{} ($\log{\mathrm{T}} \approx 6.2, 7.3$; corona and hot flare plasma) passbands, and HMI continuum images were used to provide full context of the evolution and history of each AR.

At the time of the January 2013 scan, the Solar Terrestrial Relations Observatory \citep[STEREO-A; ][]{Kaiser2008} spacecraft and SDO were located such that they provided full coverage of the Sun: SDO was located at Earth and STEREO-A (STEREO-B) was located approximately $120^\circ$ ahead of (behind) the Earth with respect to the Sun-Earth line. In the absence of magnetograms, EUV images from the Extreme Ultraviolet Imagers \citep[EUVI; ][]{Wuelser2004} in the Sun Earth Connection Coronal and Heliospheric Investigation \citep[SECCHI; ][]{Howard2008} instrument suite, particularly in the 195 \AA{} passband ($\log{\mathrm{T}} \approx 6.1, 7.2$), were used to track ARs when they were on the far side of the Sun. This allowed for a better determination of when the ARs emerged, and, therefore, a more precise age calculation for the ARs in this scan. However, the STEREO spacecraft were not available for the two scans that took place in 2015. Communications with STEREO-B were lost in 2014 and, in 2015, STEREO-A was located at an angle of approximately $180^\circ$ from the Earth which resulted in a data gap coinciding with the time running up to the EIS scans. Where the exact moment of AR emergence could not be captured (either because of spacecraft availability limiting the coverage or due to a data gap), a minimum and maximum age were determined instead. The minimum (maximum) age was given by the first available observation of the AR (last available observation before the data gap).

Synoptic map data from the Global Oscillations Network Group \citep[GONG; ][]{Harvey1996} instruments were used to identify whether an AR is part of an activity nest. These are long-lived regions of magnetic activity, where repeated flux emergence takes place. Magnetic fields brought up by each flux emergence reconnect with the preexisting field, making activity nests sites of stronger magnetic reconnection and heating rates. Where an AR emerged in a region of preexisting magnetic environment, the AR was tracked back in time for multiple rotations to identify whether repeated flux emergence took place at its location. If that was the case, both the age of the most recent significant flux emergence and the age of the nest were determined.

\subsection{FIP bias and plasma composition}
\label{EIS/composition}
FIP bias was calculated using the Hinode EIS Si {\footnotesize X} 258.38 \AA{} (low FIP, FIP = 8.15 eV) and S {\footnotesize X} 264.23 \AA{} (high FIP, FIP = 10.36 eV) ratio. The method for calculating the FIP bias in each pixel was described in detail by \citet{Brooks2011, Brooks2015} and is designed to remove temperature and density effects on the FIP bias calculation. Here, the Fe {\footnotesize XIII} 202.04 \AA{}/203.82 \AA{} ratio was used to estimate the electron density, and Fe lines Fe {\footnotesize VIII} to {\footnotesize XVI} to derive the differential emission measure (DEM). The Si {\footnotesize X}/S {\footnotesize X} diagnostic is appropriate for plasma temperatures of $\log{\mathrm{T}}\approx 6.2$, i.e. $T \approx 1.5$ MK \citep{Feldman2009b}, making it ideal for studying quiescent ARs like the ones presented in this study. The EIS study details and emission lines used are summarized in Table \ref{EIS_studies}.

\begin{deluxetable}{ll}[t]
\tabletypesize{\footnotesize}
\centering
\tablecolumns{2}
\tablehead{ \colhead{EIS Study Details}}
\tablecaption{Summary of Hinode/EIS study details and emission lines used for creating the FIP bias maps.} \label{EIS_studies}
\startdata
Study acronyms & \makecell[l]{DHB\_006 (January 2013 \& April 2015) \\ DHB\_007 (October 2015))}\\
Study numbers & \makecell[l]{491 (January 2013 \& April 2015) \\ 544 (October 2015)} \\
Emission lines used & \makecell[l]{Fe {\scriptsize X} 184.53 \AA{}, Fe {\scriptsize VIII} 185.21 \AA{}, \\ Fe {\scriptsize IX} 188.49 \AA{},  Fe {\scriptsize XI} 188. 21 \AA{}, \\ Fe {\scriptsize X} 188.29 \AA{}, Fe {\scriptsize XII} 195.12 \AA{}, \\ Fe {\scriptsize XIII} 202.04 \AA{}, Fe {\scriptsize XII} 203.72 \AA{}, \\ Fe {\scriptsize XIII} 203.82 \AA{}, Fe {\scriptsize XVI} 262.98 \AA{}, \\ Fe {\scriptsize XIV} 264.78 \AA{}, Fe {\scriptsize XV} 284.16 \AA{}, \\ Si {\scriptsize X} 258.38 \AA{}, S {\scriptsize X} 264.22 \AA{} \\ (January 2013) \\ Fe {\scriptsize X} 184.53 \AA{}, Fe {\scriptsize VIII} 185.21 \AA{}, \\ Fe {\scriptsize VIII} 186.60 \AA{}, Fe {\scriptsize IX} 188.49 \AA{}, \\ Fe {\scriptsize XII} 192.39 \AA{}, Fe {\scriptsize XI} 188. 21 \AA{}, \\ Fe {\scriptsize X} 188.29 \AA{}, Fe {\scriptsize XII} 195.12 \AA{}, \\ Fe {\scriptsize XIII} 202.04 \AA{}, Fe {\scriptsize XIII} 203.82 \AA{}, \\ Fe {\scriptsize XVI} 262.98 \AA{}, Fe {\scriptsize XIV} 264.78 \AA{}, \\ Fe {\scriptsize XV} 284.16 \AA{}, \\ Si {\scriptsize X} 258.38 \AA{}, S {\scriptsize X} 264.22 \AA{} \\ (April \& October 2015)}\\
Field of view & 492'' $\times$ 512''\\
Rastering & 2'' slit, 123 positions, 4'' coarse step\\
Exposure time & 30s\\
Total raster time & 1h 1m 30s\\
\makecell[l]{Reference spectral \\ window} & Fe {\scriptsize XII} 195.12 \AA{} \\ 
\enddata
\end{deluxetable}

\section{Method and data analysis}

\subsection {Full Sun maps}
Each full Sun map was created by stitching together 26 EIS observations (rasters) taken from 09:37 UT on the 16th to 07:06 UT on the 18th for the January 2013 scan, from 09:14 UT on the 1st to 01:49 UT on the 3rd for the April 2015 scan and from 10:27 UT on the 16th to 01:31 UT on the 18th for the October 2015 scan. Before creating the full Sun maps, pixels that had an associated $\chi^2$ value above 12 \citep[which is the number of Fe lines used for the DEM calculation, see][for a full description of the method]{Brooks2011, Brooks2015} were removed from each raster.

The individual EIS rasters are taken such that, together, they cover the entire Sun, which means that there are overlap regions with data from two or more rasters. The full Sun map value for these pixels is calculated after the FIP bias filtering, as the average of all the pixel values available for that location. The January 2013 full Sun FIP bias map (Figure \ref{FDscans}c) shows the same data that was used by \citet{Brooks2015}, but note that the display image in their Figure 3 is not directly comparable. Figure 3 in \citet{Brooks2015} shows qualitative data (the FIP bias defined as the Si {\footnotesize X} 258.38 \AA{}/S {\footnotesize X} 264.23 \AA{} intensity ratio), while Figure \ref{FDscans}c here shows the quantitative data (the FIP bias computed from the DEM calculation) - see the discussion in \citet{Brooks2015} on how their Figure 3 was constructed.

To create the corresponding full Sun magnetogram, cropped HMI images that match the start time and field of view (FOV) of each of the individual rasters are also stitched together. This is to ensure that we compare FIP bias to the magnetogram that is closest in time for each of the rasters. In the overlap regions, the most recent magnetogram data were kept.

\subsection{Active region definition}
ARs were identified using HMI line of sight (LOS) magnetic field data, and their boundaries were defined by eye in the plane of the image, tracking the evolution of the AR and looking for a sharp gradient between the magnetic flux of the AR and its surroundings. The selected contours were broad enough to include all the magnetic flux associated with the AR. This included small-scale background field between AR field fragments, but this was accounted for by filtering out pixels with an absolute magnetic flux density of less than 30 G. The same method was used for selecting the boundaries of individual polarities within an AR. In addition, when selecting the contour for one polarity, any opposite polarity field was filtered out as well.

The HMI contours were then plotted over the FIP bias map to extract the composition data. Each AR is characterised by the median FIP bias value of the distribution of values within the contour (the 50th percentile) and the spread (the difference between the 25th and 75th percentiles). Note that the spread is an indicator of the range of FIP bias values observed in the AR, rather than an error associated with the median FIP bias value. We also characterise the skewness using Kelly's skewness coefficient, $S_k = (P_{90}+P_{10}-2 \times P_{50})/(P_{90}-P_{10})$.  

\section{Results}
It has long been recognised that the strongest FIP effect is observed in ARs. ARs are sites of stronger magnetic activity, so it is likely that the magnetic field is driving the FIP effect. We aim to get a better understanding of how the magnetic field and its distribution within an AR influences the observed FIP bias. 

\subsection{FIP bias vs active region total unsigned magnetic flux and age}
\begin{figure}[t]
    \centering
    \includegraphics[width=0.45\textwidth]{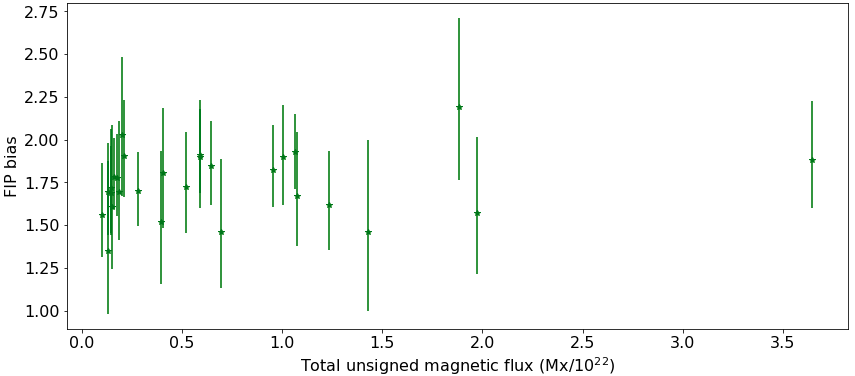}
    \caption{FIP bias variation with total unsigned magnetic flux of the active region. The vertical bars indicate the FIP bias spread, i.e. the 25th and 75th percentile of the distribution in each region. The 50th percentile is highlighted with a star.}
    \label{FIP_ARflux}
\end{figure}

\begin{figure}[t]
    \centering
    \includegraphics[width=0.45\textwidth]{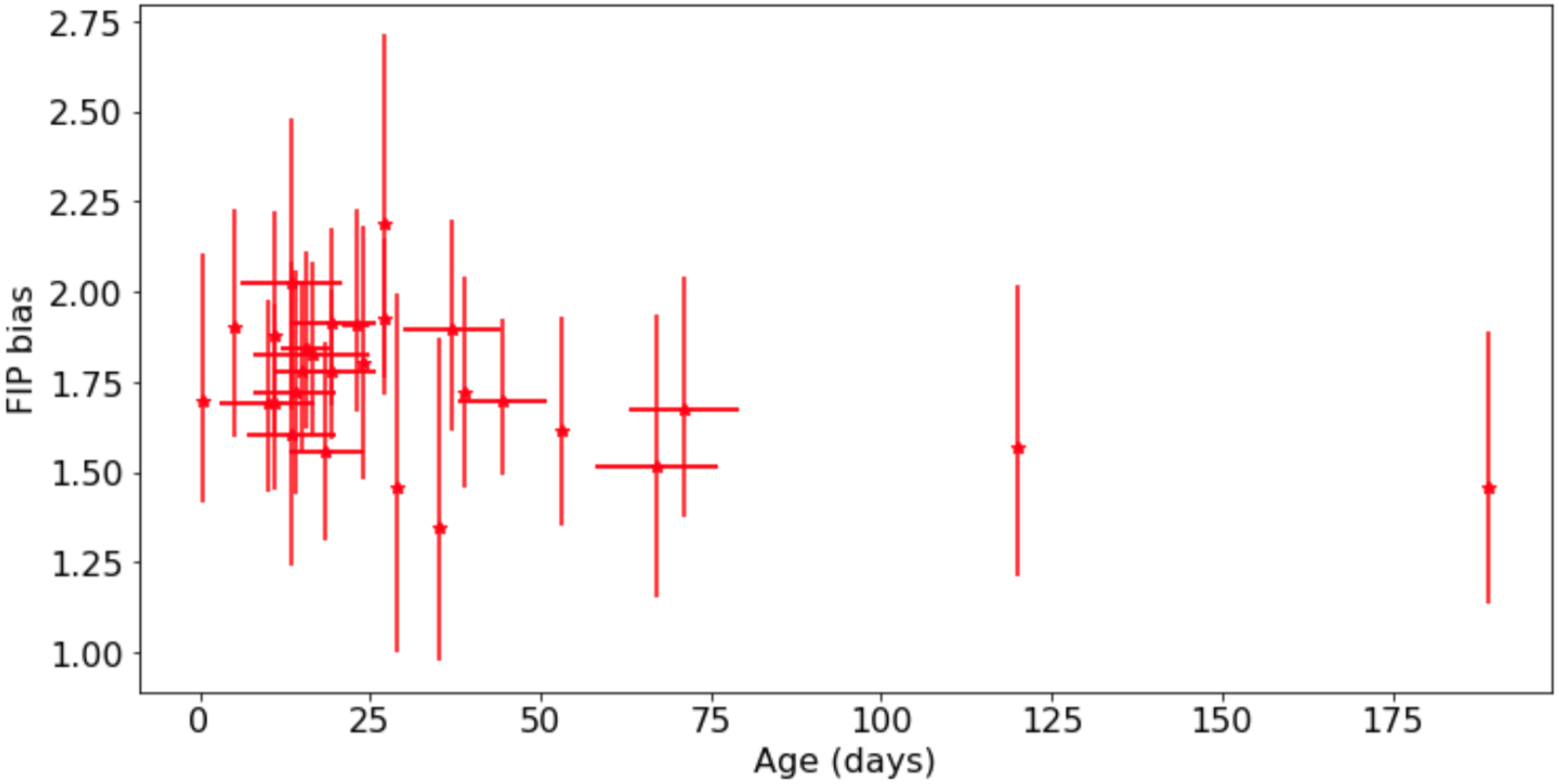}
    \caption{FIP bias variation with active region age. The vertical bars indicate the FIP bias spread, i.e. the 25th and 75th percentile of the distribution in each region. The 50th percentile is highlighted with a star. The error bar associated with the age measurement corresponds to the minimum and maximum age for that region, in case there was no observation available at the time and location of its emergence.}
    \label{FIPvsAGE}
\end{figure}

An AR's total unsigned magnetic flux varies throughout its lifetime, as a function of age and the lifetime of an AR depends on its magnetic flux content \citep{Schrijver2000}. Therefore, total unsigned magnetic flux and age must be considered together. The variation of FIP bias with AR total unsigned magnetic flux and age is shown in Figures \ref{FIP_ARflux} and \ref{FIPvsAGE}. In both plots, the vertical bars indicate the FIP bias spread, i.e. the 25th and 75th percentile of the distribution in each region, rather than a measurement error. For the ARs that emerged on a part of the Sun that was not observed by any spacecraft, the age measurement has an associated error bar that corresponds to the minimum and maximum age for that region (see Section \ref{Coronal EUV and Magnetic field observations}). The plots in Figures \ref{FIP_ARflux} and \ref{FIPvsAGE} show no global correlation between the FIP bias of an AR and total unsigned magnetic flux or age. However, this does not mean that there is no change in the FIP bias of ARs during their lifetimes. Rather, it indicates the need to study FIP bias variation in the context of the evolution of each AR and understand their individual evolutionary paths. It is likely that a normalisation of these parameters would be needed for a better comparison between ARs, i.e. normalise the total unsigned magnetic flux by peak total unsigned magnetic flux (the magnetic flux content of the AR), and normalise the age by the total lifetime of the AR. This would essentially be an indicator for the evolutionary stage of each AR, but the reduced HMI coverage throughout their lifetime would make such a calculation very difficult.

\subsection{FIP bias vs magnetic flux density}
\label{FIP bias vs magnetic flux density}
\begin{figure}[t]
    \centering
    \includegraphics[width=0.45\textwidth]{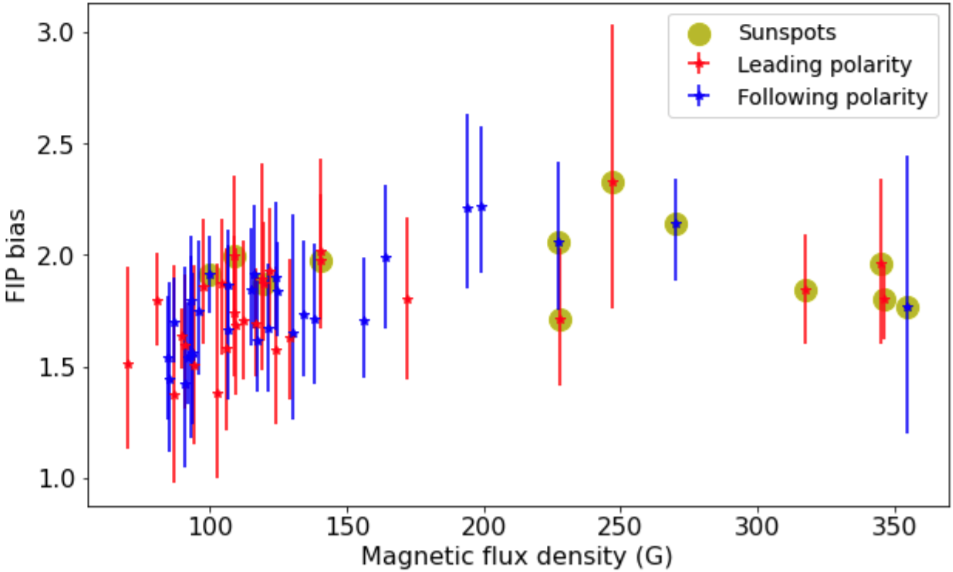}
    \caption{FIP bias variation with magnetic flux density for individual leading (red) and following (blue) polarities. The vertical bars indicate the FIP bias spread, i.e. the 25th and 75th percentile of the distribution in each region. The 50th percentile is highlighted with a star. Active regions that still have a sunspot are highlighted with a yellow dot. The Pearson correlation coefficient between the median FIP bias values and the magnetic flux density in the region $\leq200 \mathrm{~G}$ is 0.65.}
    \label{Fluxdens}
\end{figure}

As ARs evolve from emergence through decay, there is a change in their magnetic flux density. The next question we ask is whether the flux density influences the FIP bias. The variation of FIP bias as a function of magnetic flux density is shown in Figure \ref{Fluxdens}. Similar to Figures \ref{FIP_ARflux} and \ref{FIPvsAGE}, the vertical bars show the spread in FIP bias within the region. Data points in red (blue) correspond to leading (following) polarities. The yellow dots indicate polarities that still have a sunspot. For this plot, data from individual polarities were used instead of overall ARs. This decision was motivated by the asymmetries in the motion and stability of the leading and following polarities, which result in the following polarity decaying faster than the leading polarity \citep{hale1938}. Analysing them separately ensures that the magnetic flux density is more homogeneous within the selected region. In the emergence phase, the leading polarity converges immediately into a more compact and longer lived magnetic field configuration, while the following polarity may form shorter lived spots that become dispersed faster. This results in different magnetic field densities in opposite polarities, essentially placing them at different evolutionary stages. However, the asymmetry becomes weaker with time, so the separation into leading and following polarities is particularly important for younger ARs and less important for dispersed ARs.

The results show that FIP bias increases with magnetic flux density in the region $\leq200\mathrm{~G}$. The Pearson correlation coefficient between the median FIP bias values and the magnetic flux density in this region is 0.65. This is in agreement with \citet{Baker2013} who found a moderate correlation between FIP bias and magnetic flux density. It is interesting to note that, although populated with fewer data points, the same trend does not seem to continue in the region $\geq 200 \mathrm{~G}$. Above this threshold, all the data points still have sunspots.

\subsection{FIP bias in leading vs. following polarities}

\begin{figure}[t]
    \centering
    \includegraphics[width=0.45\textwidth]{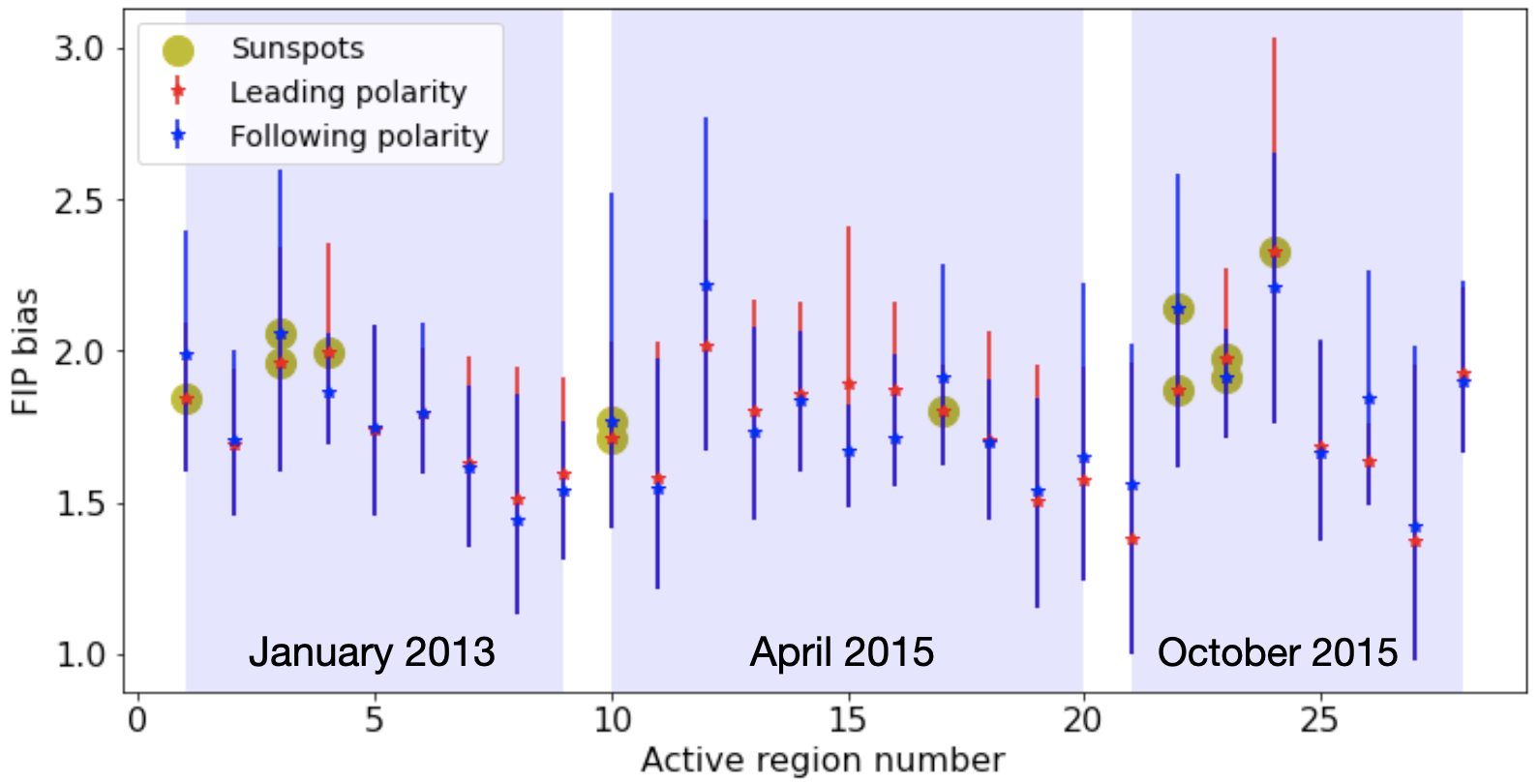}
    \caption{FIP bias in leading (red) vs. following (blue) polarities. The x axis indicates the active region R code (1-28) used within the dataset, as defined in Figure \ref{FDscans}. The vertical bars indicate the FIP bias spread, i.e. the 25th and 75th percentile of the distribution in each region. The 50th percentile is highlighted with a star. Active regions that still have a sunspot are highlighted with a yellow dot.}
    \label{LPvsFP}
\end{figure}

A comparison between the median FIP bias values in the leading vs. following polarities of the ARs is shown in Figure \ref{LPvsFP}. The plot shows that 10 ARs (35\% of the sample) have higher FIP bias in the following polarity, 8 ARs (28\%) have higher FIP bias in the leading polarity and the remaining 10 ARs (35\%) have approximately the same FIP bias in both polarities (difference smaller than 0.05 in FIP bias). The higher FIP bias values registered in one polarity or another are not correlated with the AR's position on the disc, indicating that this difference is due to an asymmetry between the two polarities rather than a projection effect. 

Across the dataset, there is a large variety of ARs at different evolutionary stages: an emerging AR (R10), very decayed ARs that have formed filament channels along their main PILs (e.g., R8, R11, R21), ARs that have compact magnetic field in both polarities (e.g., R3, R10, R24) or a single polarity (e.g., R1), ARs that have decayed far beyond having homogeneous field in either the leading or the following polarities (e.g., R7, R14, R19), ARs that are dominated by one polarity (e.g., R8). It is likely that this very varied coronal field configuration is the reason why no systematic trend is seen. 

\begin{figure*}[t]
    \centering
    \includegraphics[width=1.0\textwidth]{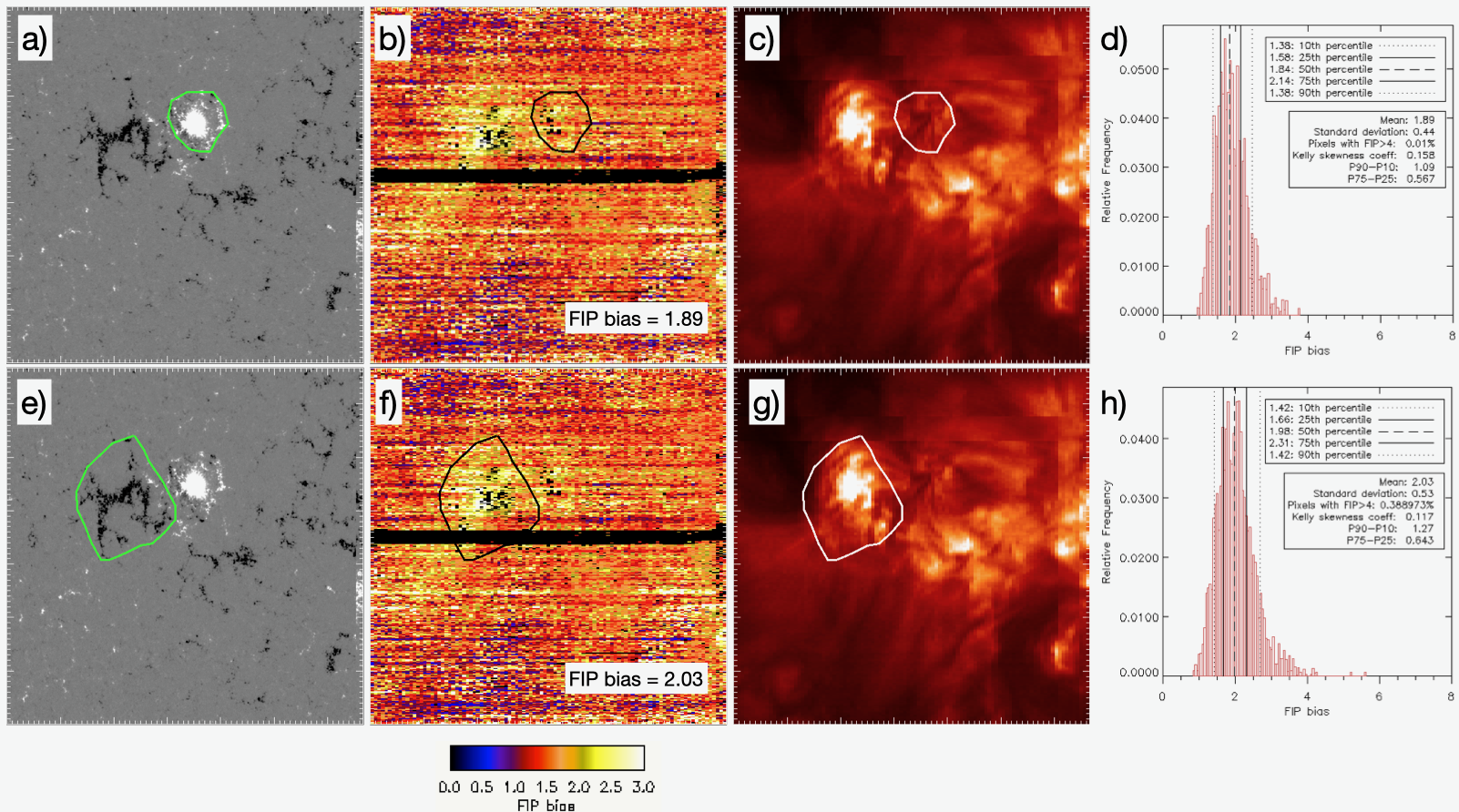}
    \caption{Case study of the median FIP bias in the leading (top) vs. following (bottom) polarities of R1 (January 2013). The maps show a,e) HMI LOS photospheric magnetic field strength; b, f) Hinode/EIS FIP bias; c, g) Hinode/EIS Fe {\footnotesize XIII} 202.04 \AA{} intensity; d,h) histograms of the FIP bias within the contours shown in green, black, and white on the other panels. The values in the two boxes show the median FIP bias value in the defined contour.}
    \label{LPvsFPcasestudy}
\end{figure*}

A case study of R1 (January 2013) is shown in Figure \ref{LPvsFPcasestudy}. The leading polarity has lower FIP bias than the following polarity. This AR is an example showing the asymmetric evolution of the magnetic field in the leading polarity as compared to the following. The leading polarity is more compact and still has a sunspot, while the following polarity is already dispersed. The asymmetry is also reflected in the different FIP bias values, likely due to the fact that the higher flux density above the sunspot is actually decreasing the overall FIP bias (see Section \ref{FIP bias vs magnetic flux density}). This AR is located close to disc centre, such that differences in FIP bias are likely due to asymmetries in the opposite polarities rather than projection effects.

\subsection{Regions at different evolutionary stages}
\label{Regions at different evolutionary stages}
The ARs in the dataset are at different stages in their evolution and can, therefore, offer insight into FIP bias values in these different stages. For this, we categorise the ARs into 5 groups based on their evolutionary stages (see Table \ref{EvolutionaryStage}). The lifetime of an AR is typically divided into emergence phase and decay phase. The decay phase is always longer than the emergence phase, but it varies from around 70\% of the total lifetime for ephemeral ARs to as much as 97\% for large ARs \citep{VanDriel-Gesztelyi2015a}. As the decay phase is so much longer, here we divide it even further into substages: ARs that still have sunspots (at the peak development or in their early decay phase), decayed ARs (sunspots have disappeared), extended and very dispersed ARs (field is so dispersed that it is not easily distinguished from the quiet Sun), and ARs with filament channels.

The results shown in Table \ref{EvolutionaryStage} suggest that, generally, the FIP bias of fully developed ARs is higher than that of the emerging AR and then it reduces for the progressively more dispersed groups. \citet{Baker2018} found a dependence of FIP bias on AR evolution for seven emerging flux regions. The present result indicates that the same behaviour is found in small and large ARs as well. 

The lowest FIP bias values are found in ARs that formed filament channels. A case study example of AR R11 is shown in Figure \ref{R11}. The filament channel is seen as the dark feature in the EUV emission (Figure \ref{R11}a). This corresponds to a corridor of low magnetic field strength along the PIL in the associated magnetogram (Figure \ref{R11}b), which is a sign of ongoing flux cancellation taking place along the PIL to form the filament channel structure. The FIP bias map (Figure \ref{R11}c) indicates that the filament channel has distinctly lower FIP bias values than the rest of the AR. It is likely that this is due to flux cancellation taking place in the lower atmosphere \citep{Baker2022}. Post reconnection loops bring photospheric material up into the corona, and plasma mixing leads to an overall lower FIP bias value.

\subsection{FIP bias distribution}
FIP bias within the ARs has a significant spread (difference between the 75th and 25th percentiles of the FIP bias distribution), which varies between 0.4 and 1.0. The spread values for all the ARs are given in Table \ref{ARcharacteristics}, and individual AR contours and characteristics are shown in Appendix \ref{IndividualContours}. Reference examples for the distribution of  FIP bias values in a quiet Sun and coronal hole regions are shown in Figure \ref{QSCHcontours} of the Appendix.

\begin{figure*}[t]
    \centering
    \includegraphics[width=1.0\textwidth]{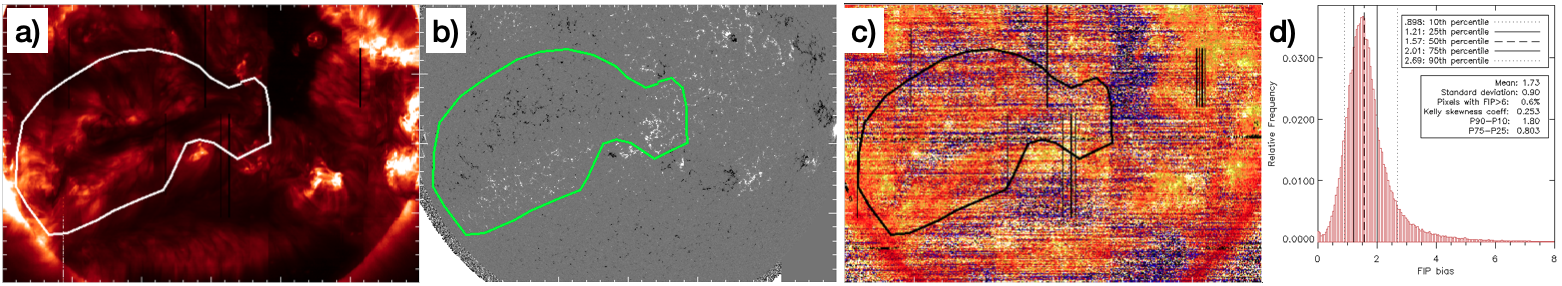}
    \caption{Case study of R11 (April 2015): a) Hinode/EIS Fe {\footnotesize XIII} 202.04 \AA{} intensity, b) HMI line of sight magnetic field strength, c) Hinode EIS FIP bias, d) FIP bias distribution within the active region contour.}
    \label{R11}
\end{figure*}

\begin{deluxetable*}{lcc}
\centering
\tablecolumns{5}
\tablecaption{Median FIP bias and Kelly's skewness coefficient ranges for the active regions in each category, excluding those active regions that are located close to the limb (outside $\pm 60^\circ$ longitude). \label{EvolutionaryStage}}
\tablehead{\colhead{Evolutionary stage (no. of ARs)} & \colhead{Median FIP bias range} & \colhead{Kelly's skewness coefficient range}}
\startdata
Emerging active regions (1) & 1.7 & 0.16 \\
Active regions with spots (4) & 1.9 - 2.2 & 0.10 - 0.23 \\
Decayed active regions (8) & 1.6 - 2.0 & 0.13 - 0.27 \\
Extended, very dispersed active regions (6) & 1.4 - 1.8 & 0.09 - 0.34 \\
Active regions with filament channels (4) & 1.5 - 1.6 & 0.23 - 0.25 \\
\enddata
\end{deluxetable*}

This spread is likely an indicator of different substructures within an AR having different FIP bias values. A case study example of R11 is shown in Figure \ref{R11}. This is a very decayed AR, with lower FIP bias values in the filament channel along its main PIL and higher FIP bias values in the arcade loops and hotter areas, which is likely why this AR has a relatively high FIP bias spread. 

The lowest spread is seen in ARs that are located close to the limb, which is likely due to the method used for defining the AR contours. The HMI magnetic field is a photospheric measurement, while the FIP bias is a coronal observation. Magnetic field expands into the corona compared to the photosphere, so using a photospheric magnetic field contour will likely introduce a challenge in fully capturing the coronal loops. Closer to the limb, this effect is amplified by projection effects.

Additionally, all distributions were found to be right hand skewed (see Table \ref{ARcharacteristics}). The skewness coefficient has a wide range of values for all the evolutionary stages between ARs with spots and very dispersed ARs (see Table \ref{EvolutionaryStage}). However, in the ARs with filament channels category, the skewness coefficient values do not exhibit this variation.

\section{Discussion}
By taking the median FIP bias value (50th percentile value) to be representative of each AR (see Table \ref{ARcharacteristics} and Figures \ref{IndividualARContours1}-\ref{IndividualARContours6} in the Appendix), we find that the median FIP bias values fall in the range 1.4 to 2.2, very similar to the values found by \citet{Baker2018} which varied between 1.2 and 2.0. \citet{Baker2018} analysed emerging flux regions with a total unsigned magnetic flux of $0.13-38 \times 10^{20}$ Mx, while the ARs in the present study range from small to large, and have total unsigned magnetic flux values of $1-36.4 \times 10^{21}$ Mx. Very similar FIP bias values are observed, in spite of the significant difference in total magnetic flux, which is a further indication that FIP bias is not influenced by the total magnetic flux. For reference, the median FIP bias values for a representative example of quiet Sun was 1.5 and for coronal hole 1.0 (see Figures \ref{IndividualARContours1}-\ref{IndividualARContours6} in the Appendix)

Magnetic flux density, however, appears to play a role. FIP bias increases with magnetic flux density in the region $\leq 200 \mathrm{~G}$, but that trend appears to stop for the data points $\geq 200 \mathrm{~G}$, which all belong to regions with sunspots. In the region $\leq 200 \mathrm{~G}$, increased magnetic flux density drives stronger heating at the chromospheric loop footpoints, ionising higher proportion of elements and, therefore, driving a stronger fractionation process which increases the FIP bias. In contrast, in the region $\geq 200 \mathrm{~G}$ the strong magnetic field concentration in the umbra of sunspots can inhibit convection and lower the temperature at chromospheric level which means a lower proportion of elements are being ionized, thus producing a lower FIP effect. This scenario is supported by the study of \citet{Baker2021} who found that the FIP bias in the umbra of a very strong sunspot has photospheric values.

Also, it is interesting to note that an AR moves from higher to lower magnetic flux density throughout its evolution (so from right to left on the plot in Figure \ref{Fluxdens}). Most of the ARs in the dataset are in different stages of the decay phase. The ones with a flux density $\geq 200 \mathrm{~G}$ still have sunspots, and they are in the early decay phase, while the ones $\leq 100 \mathrm{~G}$ are in the late decay phase. The trend of FIP bias decreasing with decreasing magnetic flux density, is, therefore, an indirect indication that FIP bias decreases with time in the AR decay phase. This result is in agreement with the previous result of \citet{baker2015a} who found that, in the decay phase of an AR, FIP bias is decreasing and remains coronal for a longer time only in a part of the AR's high flux density core. 

In the flux density region of $\leq 200 \mathrm{~G}$, FIP bias is decreasing from 2.2 to 1.4, where the 1.4 is observed in two ARs that are very dispersed to the point where they are hard to distinguish from quiet Sun. This is in line with the results of \citet{Ko2016}, who found that, in a decaying AR, FIP bias decreases from 1.8 over 3 days until it settles at a value of 1.5, which they describe as a ‘basal’ state of the quiet Sun. The FIP bias method and line ratio used in their study is very similar to the one used in this study, which means the values can be compared directly.

The FIP bias distribution within the ARs has a significant spread, which indicates that there is a range of physical processes in different AR substructures that influence the FIP bias in different ways. An interesting example is presented by the four regions that have formed filament channels along their main PIL. While having the lowest overall FIP bias values, they show a high spread and percentage of high FIP bias values. The spatial distribution of the FIP bias indicates a closer to photospheric value in the filament channel with higher FIP bias values surrounding the channel in the remnant AR arcade field. Having these substructures with different plasma composition increases the spread of the FIP bias distribution. 

The lack of general trends of FIP bias with total magnetic flux and age or systematic reasons for the observed differences in the FIP bias in leading and following polarities, as well as the dependence of FIP bias on the AR evolutionary stage, indicate that the processes influencing the composition of an AR are complex and specific to its evolution, history and magnetic configuration or environment. 

It is interesting to compare the AR FIP bias values to in situ studies of slow solar wind composition. Although the slow solar wind is believed to originate from ARs, the in situ composition measurements find higher FIP bias values than the ones observed in the presented ARs. \citet{vonSteiger2000} found that the average FIP bias of the slow solar wind, relative to O and averaged over three low-FIP elements (Mg, Si, Fe) is 2.6. This is quite high, compared to the median FIP bias values presented in this study (1.4 to 2.2). \citet{Brooks2015} analysed the January 2013 full Sun scan and identified potential slow solar wind sources at the edges of a number of ARs that were present on the surface of the Sun at the time. Notably, AR11654 (here R3) showed strong upflows on its eastern edge and was further investigated by \citet{stansby2020c}. The FIP bias values observed both in the upflow region (remotely, using EIS data) and in the solar wind (in situ, using ACE data) are generally higher than the median FIP bias for that region. It is possible that the higher FIP bias values in the upflow regions contribute to the skewed part of the distributions presented here. \citet{Brooks2011} also found that AR upflows can be a source of slow solar wind, with the EIS FIP bias values in the region of interest being within 20\% of the ACE in situ measurements.

\section{Conclusions}
In this study, we investigate the characteristics of coronal plasma elemental composition across 28 ARs of varying size, magnetic complexity and age. The sample of ARs includes one in its emergence phase, with the longest lived in the study being 244 days old. Plasma composition is determined through the FIP bias value, which indicates by how much low-FIP elements are enhanced in the corona relative to high-FIP elements, and the range of ARs studied enables an analysis of how plasma composition might be influenced by magnetic field strength, age, complexity and evolutionary stage. 

OUr findings show that there appears to be no correlation between FIP bias and total flux content of an AR or its age, which highlights our overall conclusion that plasma composition is affected by characteristics of the region that relate to its specific evolutionary journey. Our study finds that young ARs have closer to photospheric composition, and the FIP bias then increases in the ARs that are more developed and formed spots. The FIP bias then decreases in the progressively more decayed ARs, with the lowest values being observed in ARs that are very dispersed and formed filament channels along their PILs. The FIP bias dependence on the evolutionary stage of the AR is also supported by the trend of FIP bias decreasing with magnetic flux density in the region $\leq 200 \mathrm{~G}$. This is an indirect indication that FIP bias decreases with time in the AR decay phase, in agreement with previous findings \citep{baker2015a}.

The median FIP bias values found in these ARs are generally lower than the FIP bias values observed in situ in the slow solar wind \citep{vonSteiger2000, stansby2020c}. This could suggest that the slow solar wind originates from the part of an AR that has stronger FIP bias, emphasising the importance of understanding physical processes at play in these locations.

\begin{acknowledgements}

We thank Pascal Demoulin for insightful discussions and inspiration of the project. 
Hinode is a Japanese mission developed and launched by ISAS/JAXA, collaborating with NAOJ as a domestic partner, and NASA and STFC (UK) as international partners. 
Scientific operation of Hinode is performed by the Hinode science team organized at ISAS/JAXA. 
This team mainly consists of scientists from institutes in the partner countries. 
Support for the post-launch operation is provided by JAXA and NAOJ (Japan), STFC (UK), NASA, ESA, and NSC (Norway). 
T.M. acknowledges the STFC PhD studentship grant ST/V507155/1.
D.B. is funded under STFC consolidated grant number ST/S000240/1 and L.v.D.G. is partially funded under the same grant.
L.v.D.G. acknowledges the Hungarian National Research, Development and Innovation Office grant OTKA K-131508.
D.M.L. is grateful to the Science Technology and Facilities Council for the award of an Ernest Rutherford Fellowship (ST/R003246/1).
The work of D.H.B. was performed under contract to the Naval Research Laboratory and was funded by the NASA Hinode program.

\end{acknowledgements}

\newpage
\bibliography{FDproject, name}
\bibliographystyle{apalike}

\newpage
\appendix
\section{Selected contours and distributions for all the regions considered}
\label{IndividualContours}

\begin{figure}[h!]
    \centering
    \includegraphics[width=1.0\textwidth]{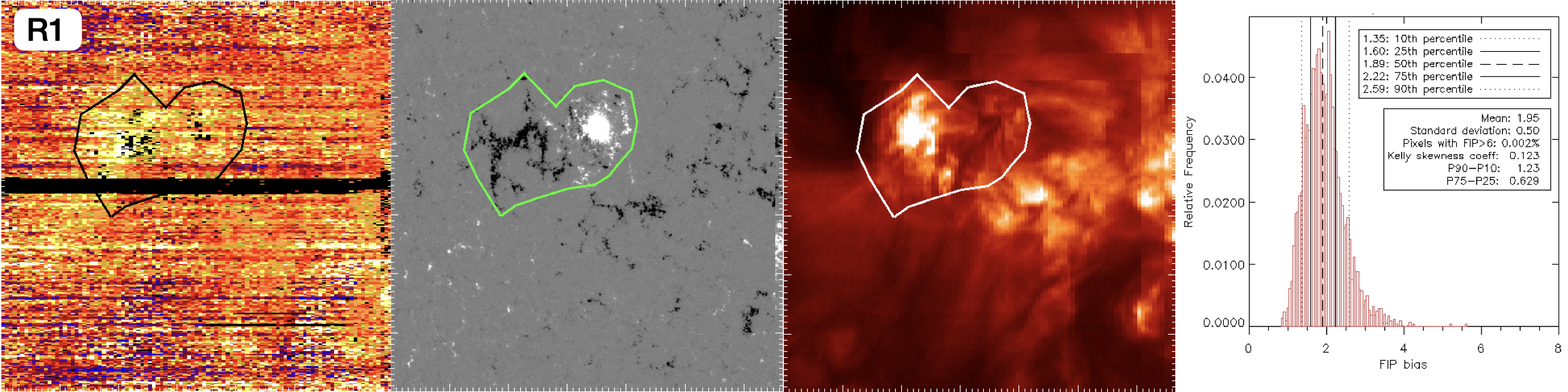}
    \includegraphics[width=1.0\textwidth]{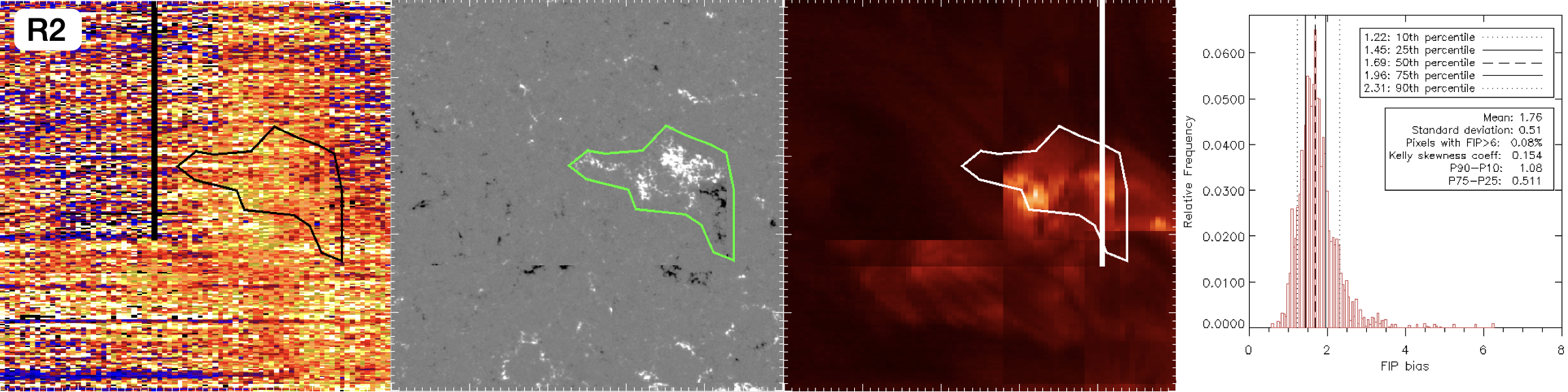}
    \includegraphics[width=1.0\textwidth]{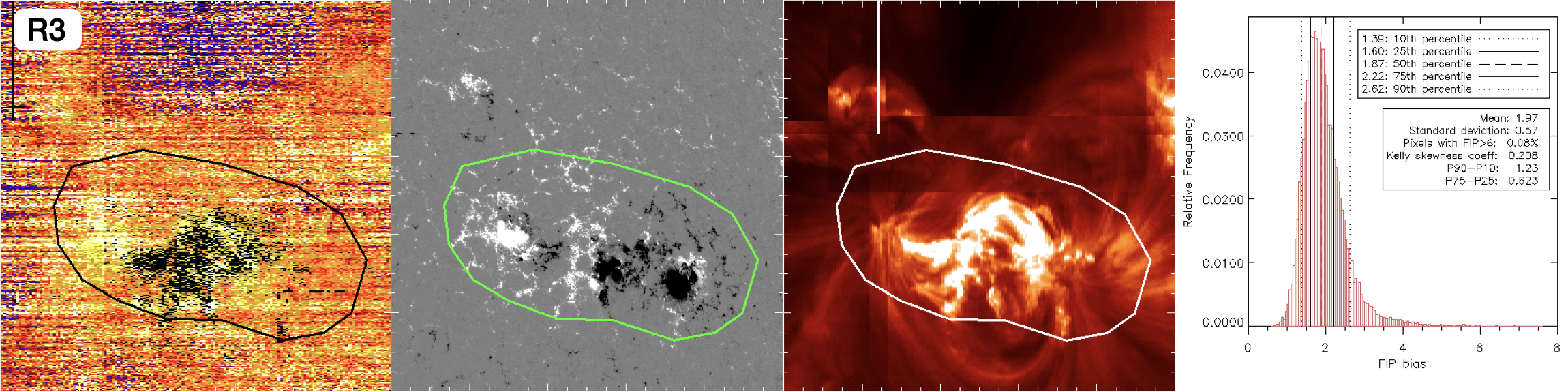}
    \includegraphics[width=1.0\textwidth]{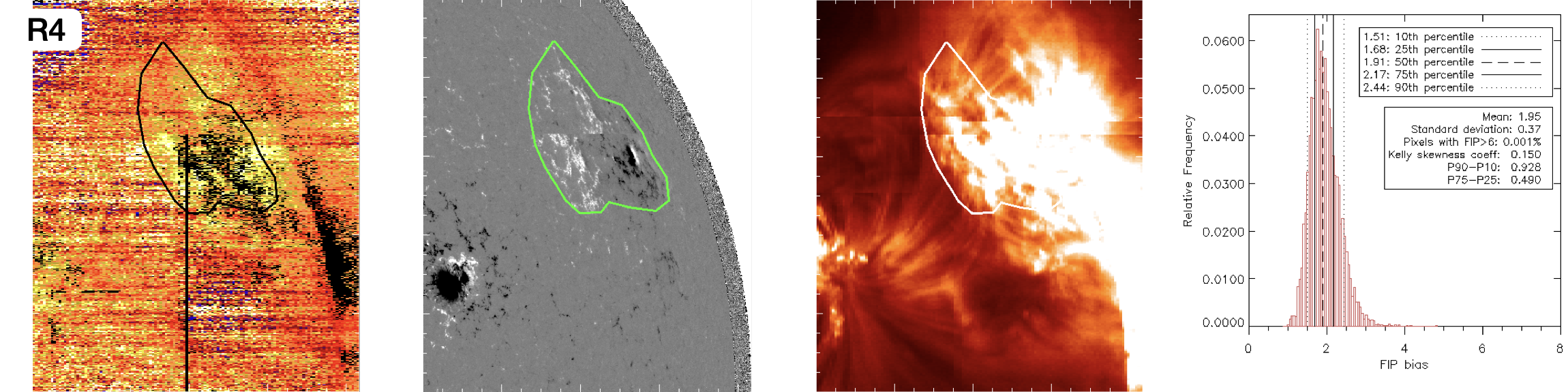}
    \caption{FIP bias distributions in the selected active region contours indicating the magnetic field boundaries of the active regions. From left to right: Hinode EIS FIP bias, HMI line of sight magnetic field strength, Hinode/EIS Fe {\footnotesize XIII} 202.04 \AA{} intensity, and FIP bias distribution within the active region contour (shwon in black, green and white respectively).}
    \label{IndividualARContours1}
\end{figure}

\begin{figure}[]
    \centering
    \includegraphics[width=1.0\textwidth]{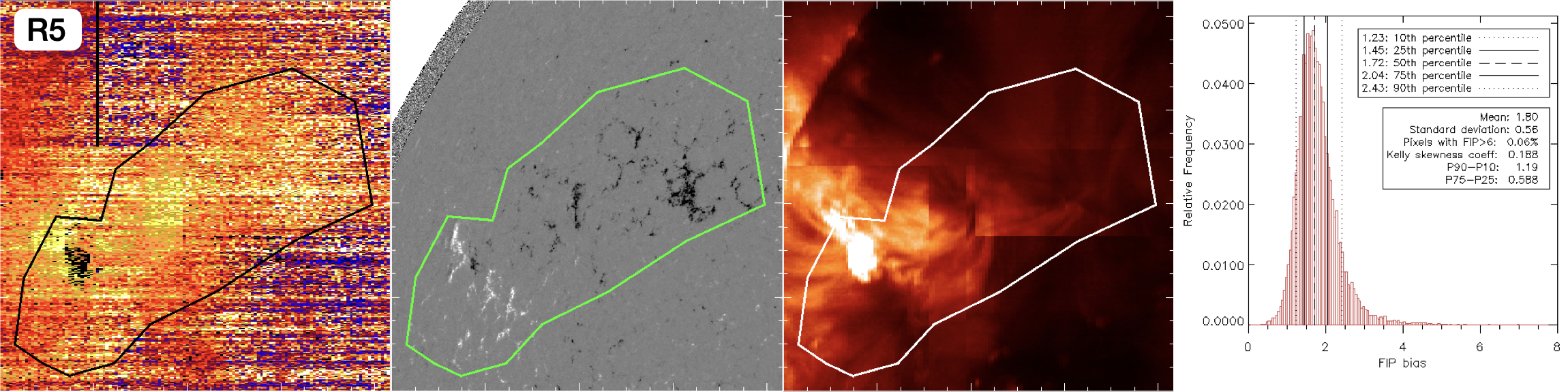}
    \includegraphics[width=1.0\textwidth]{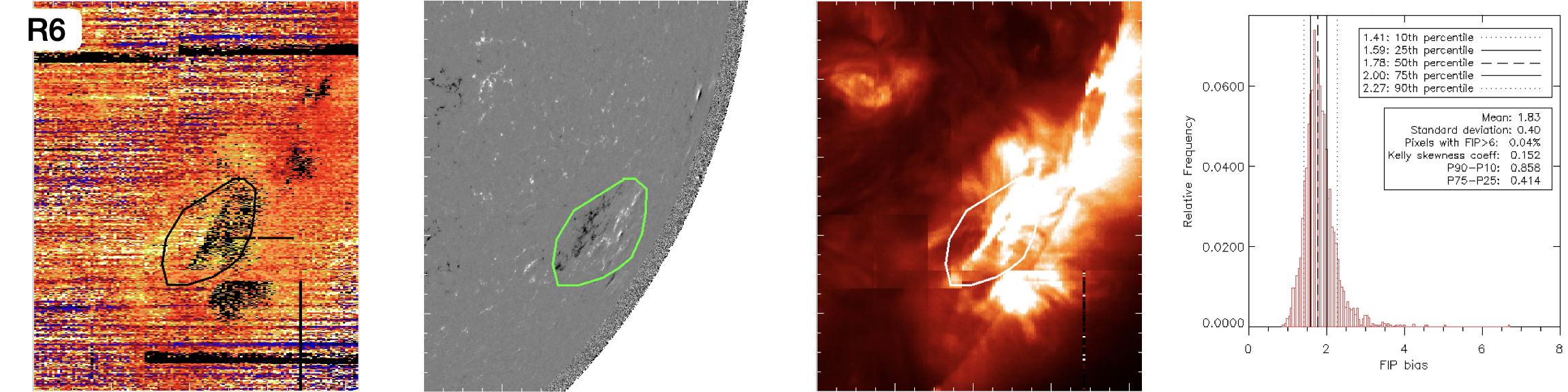}
    \includegraphics[width=1.0\textwidth]{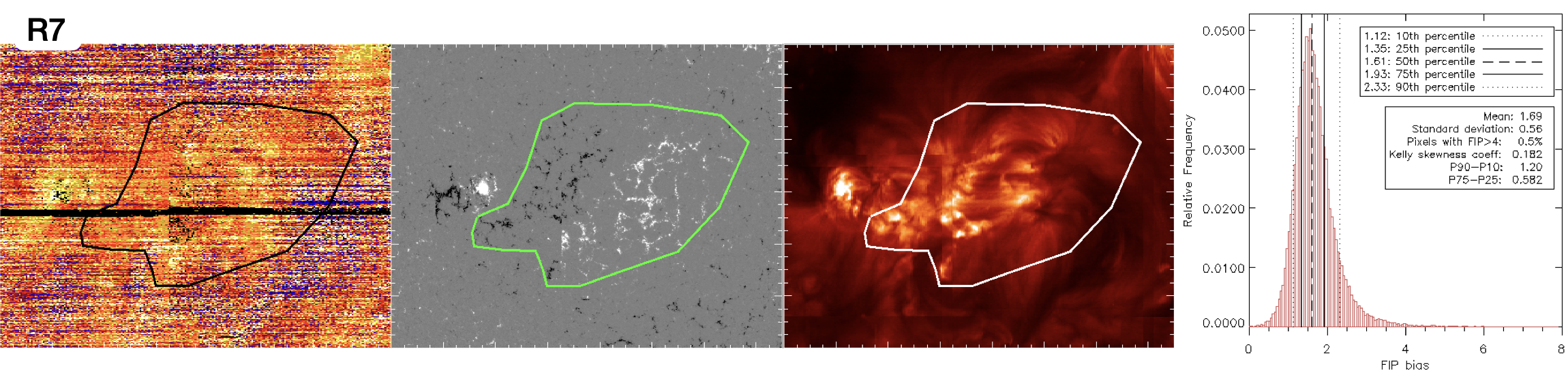}
    \includegraphics[width=1.0\textwidth]{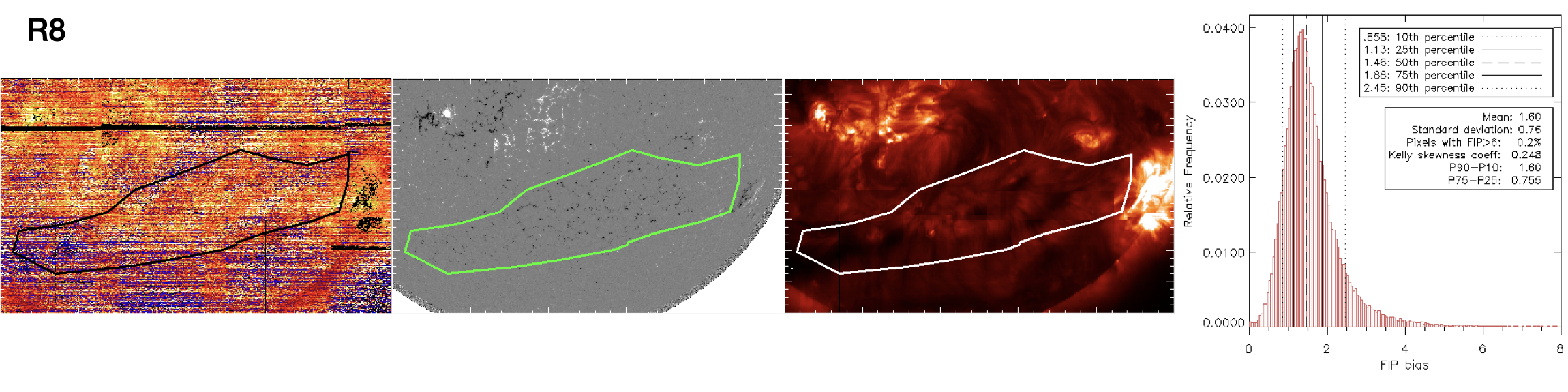}
    \includegraphics[width=1.0\textwidth]{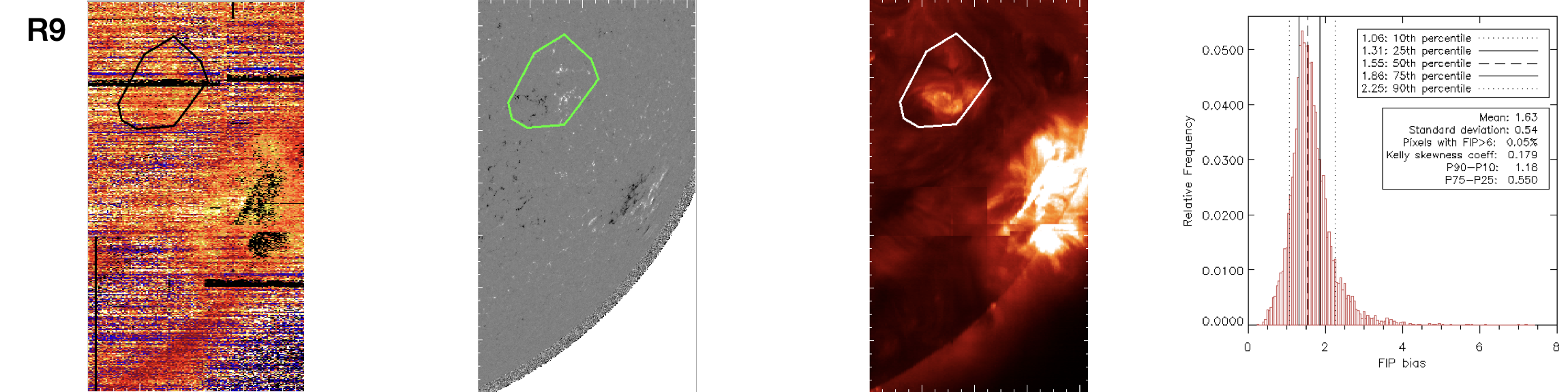}
    \caption{FIP bias distributions in the selected active region contours. From left to right: Hinode EIS FIP bias, HMI line of sight magnetic field strength, Hinode/EIS Fe {\footnotesize XIII} 202.04 \AA{} intensity, and FIP bias distribution within the active region contour (shown in black, green and white respectively).}
    \label{IndividualARContours2}
\end{figure}

\begin{figure}[]
    \centering
    \includegraphics[width=1.0\textwidth]{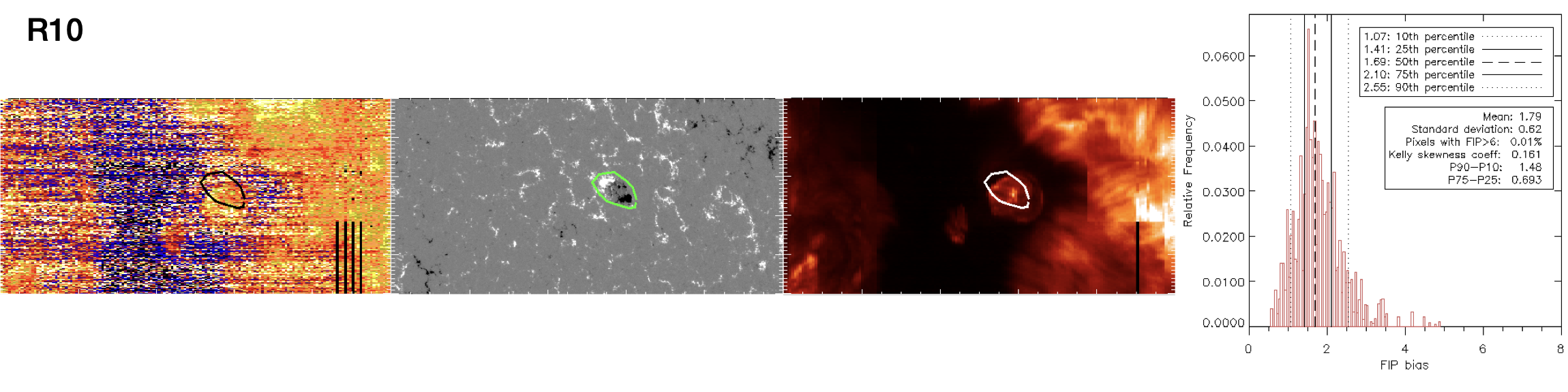}
    \includegraphics[width=1.0\textwidth]{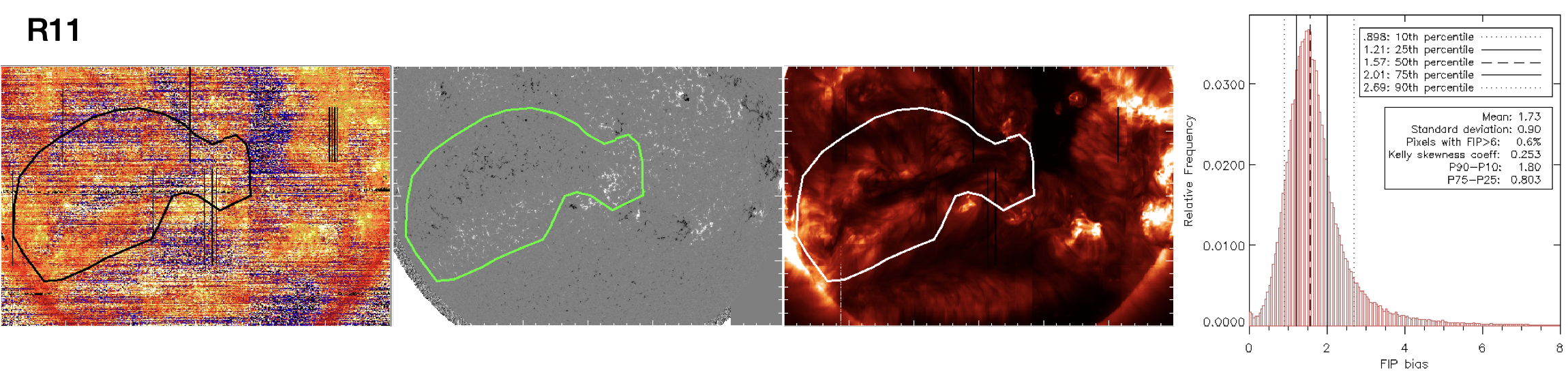}
    \includegraphics[width=1.0\textwidth]{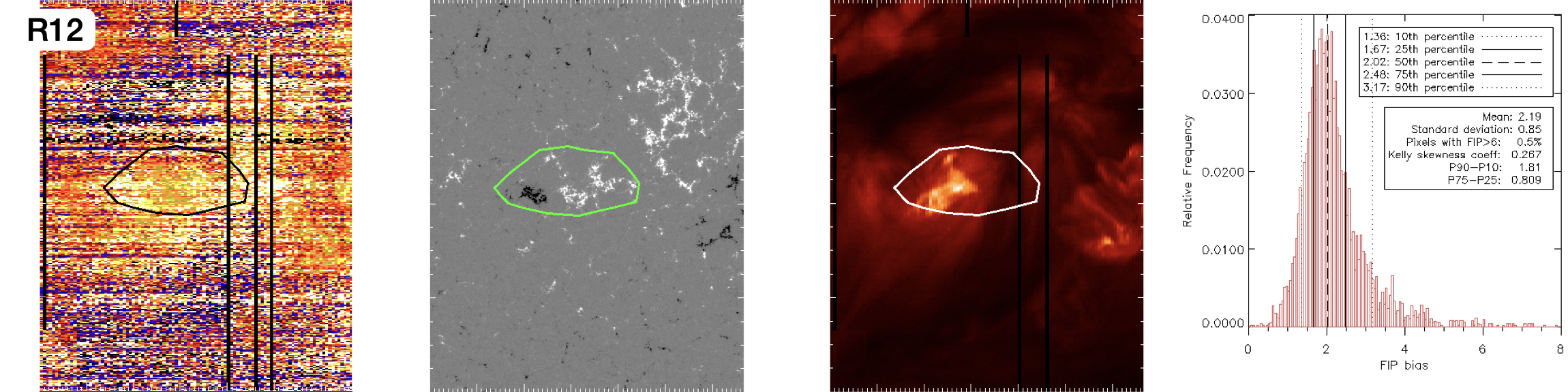}
    \includegraphics[width=1.0\textwidth]{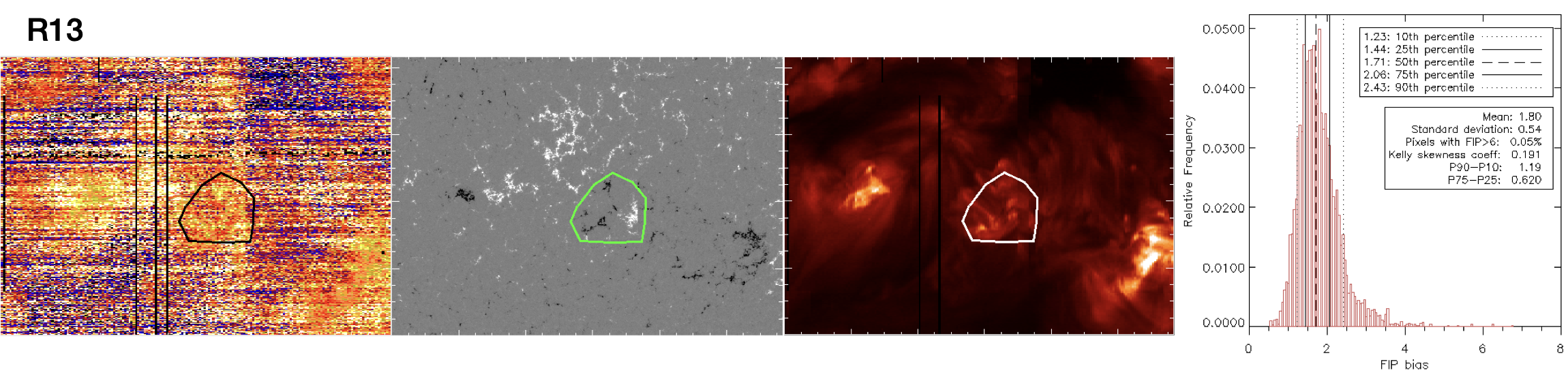}
    \includegraphics[width=1.0\textwidth]{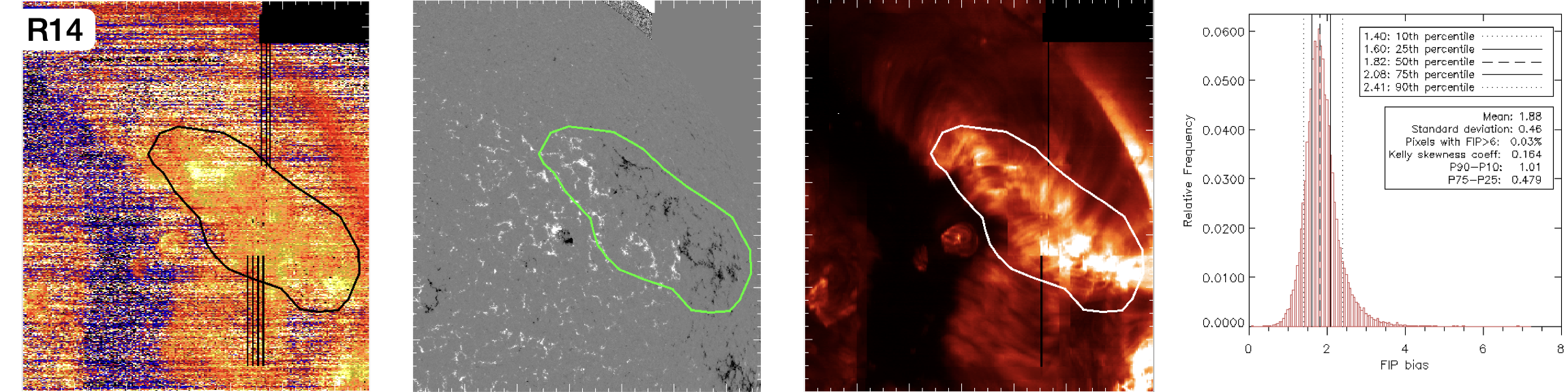}
    \caption{FIP bias distributions in the selected active region contours. From left to right: Hinode EIS FIP bias, HMI line of sight magnetic field strength, Hinode/EIS Fe {\footnotesize XIII} 202.04 \AA{} intensity, and FIP bias distribution within the active region contour (shown in black, green and white respectively).}
    \label{IndividualARContours3}
\end{figure}

\begin{figure}[]
    \centering
    \includegraphics[width=1.0\textwidth]{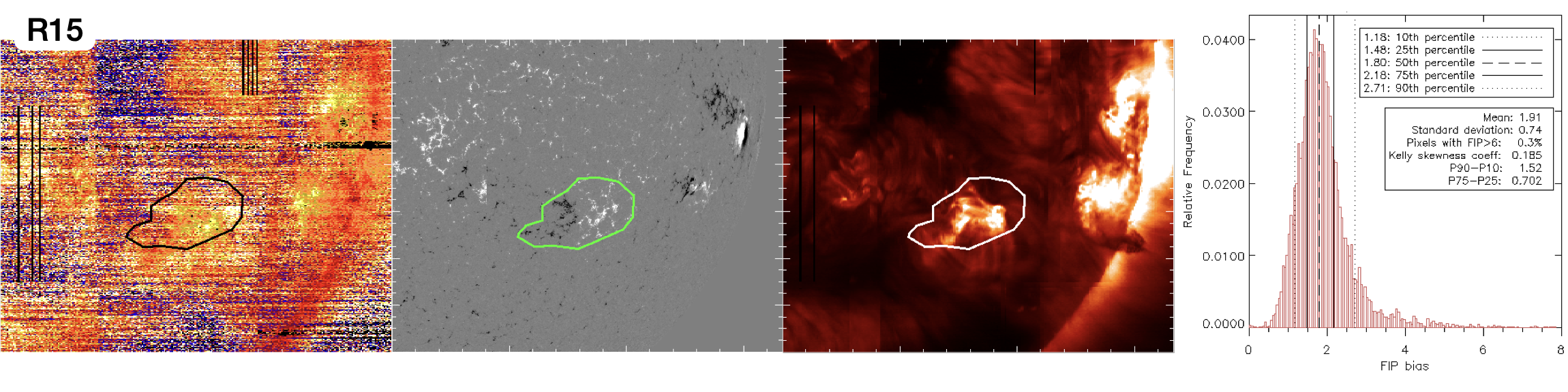}
    \includegraphics[width=1.0\textwidth]{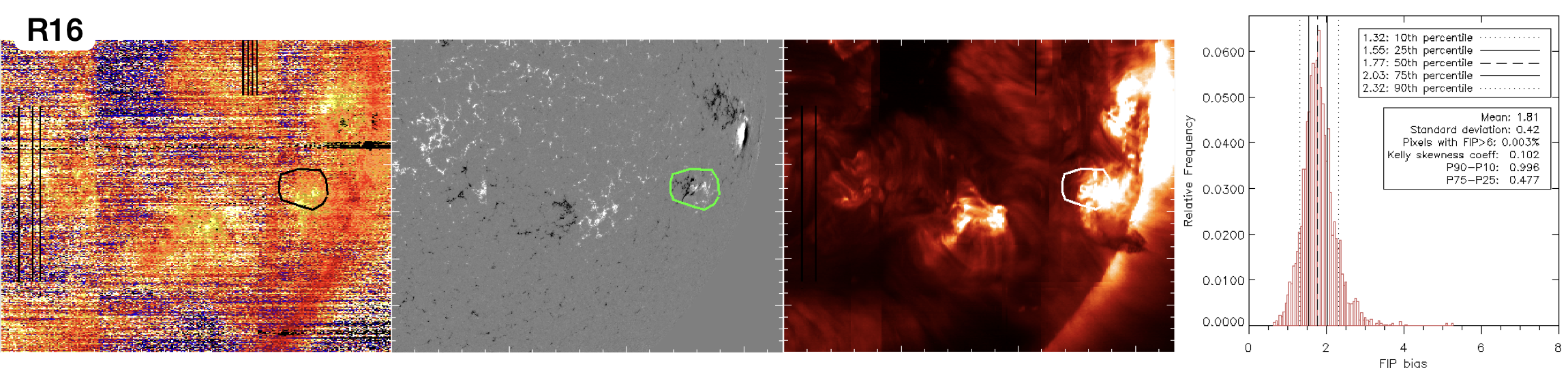}
    \includegraphics[width=1.0\textwidth]{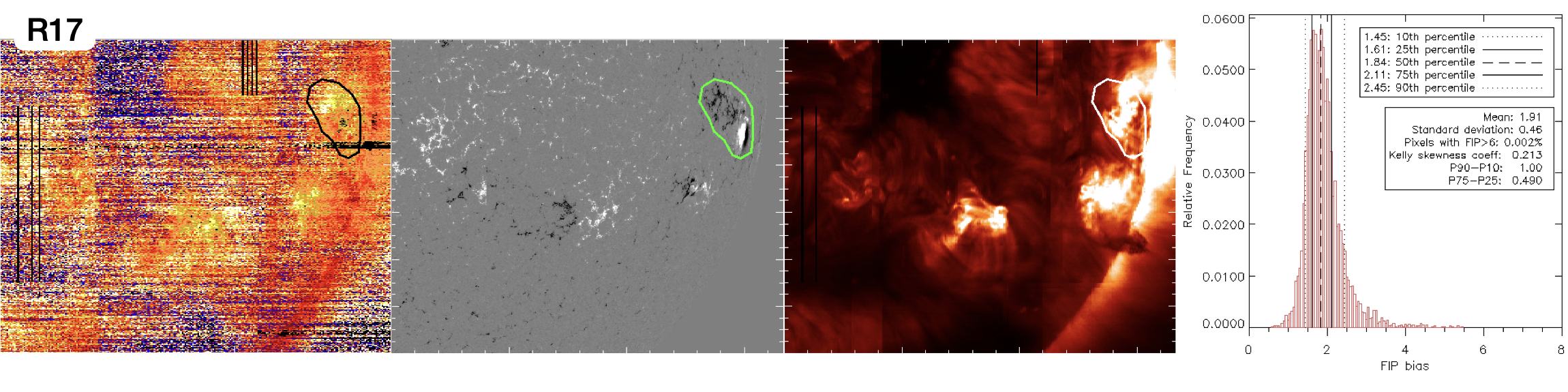}
    \includegraphics[width=1.0\textwidth]{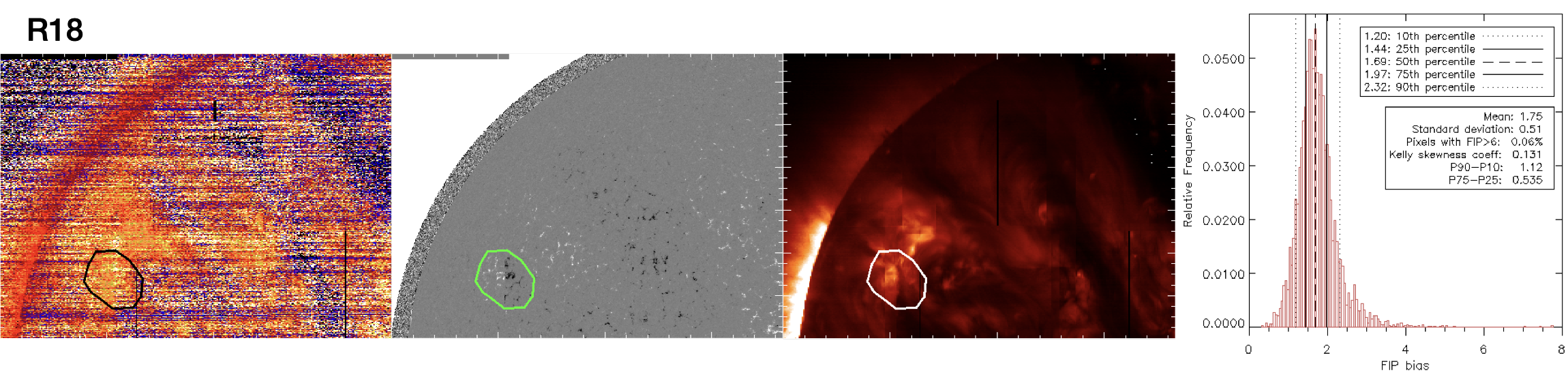}
    \includegraphics[width=1.0\textwidth]{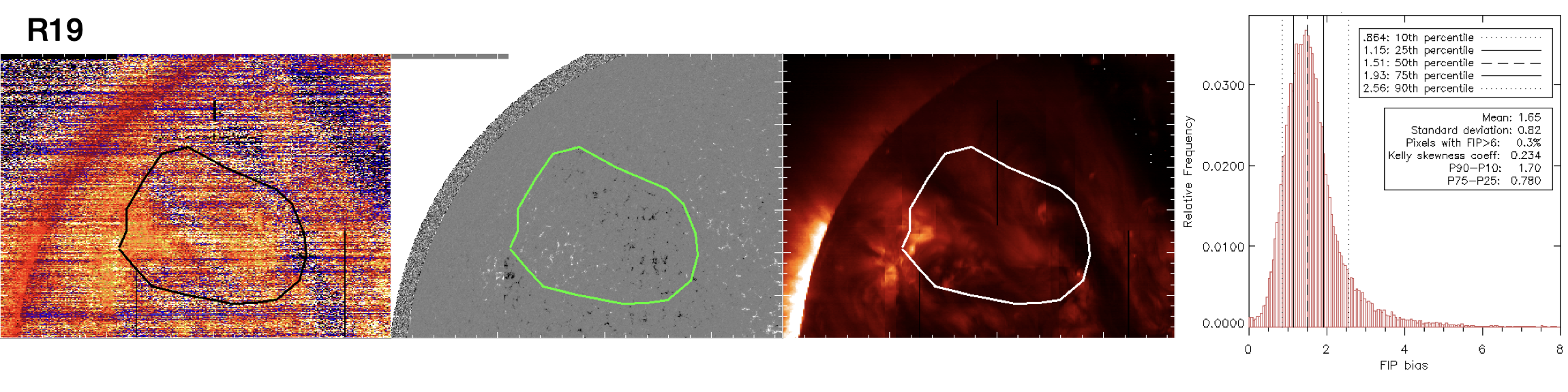}
    \caption{FIP bias distributions in the selected active region contours. From left to right: Hinode EIS FIP bias, HMI line of sight magnetic field strength, Hinode/EIS Fe {\footnotesize XIII} 202.04 \AA{} intensity, and FIP bias distribution within the active region contour (shown in black, green and white respectively).}
    \label{IndividualARContours4}
\end{figure}

\begin{figure}[]
    \centering
    \includegraphics[width=1.0\textwidth]{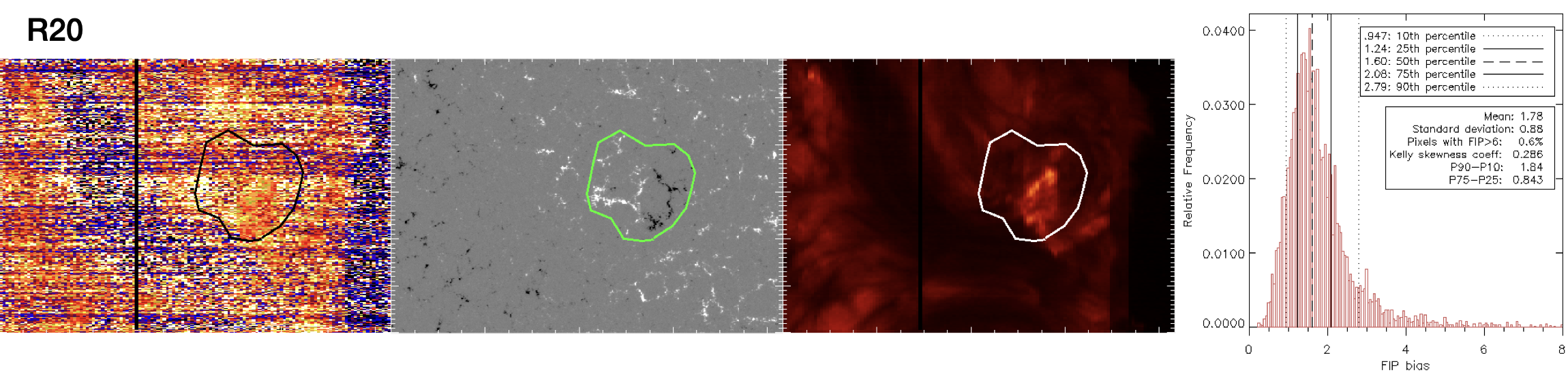}
    \includegraphics[width=1.0\textwidth]{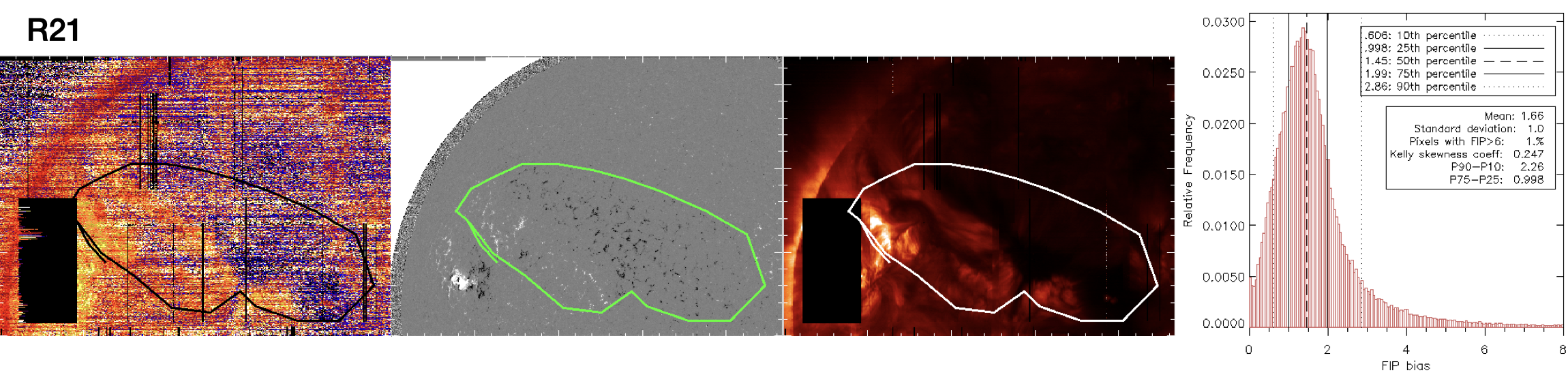}
    \includegraphics[width=1.0\textwidth]{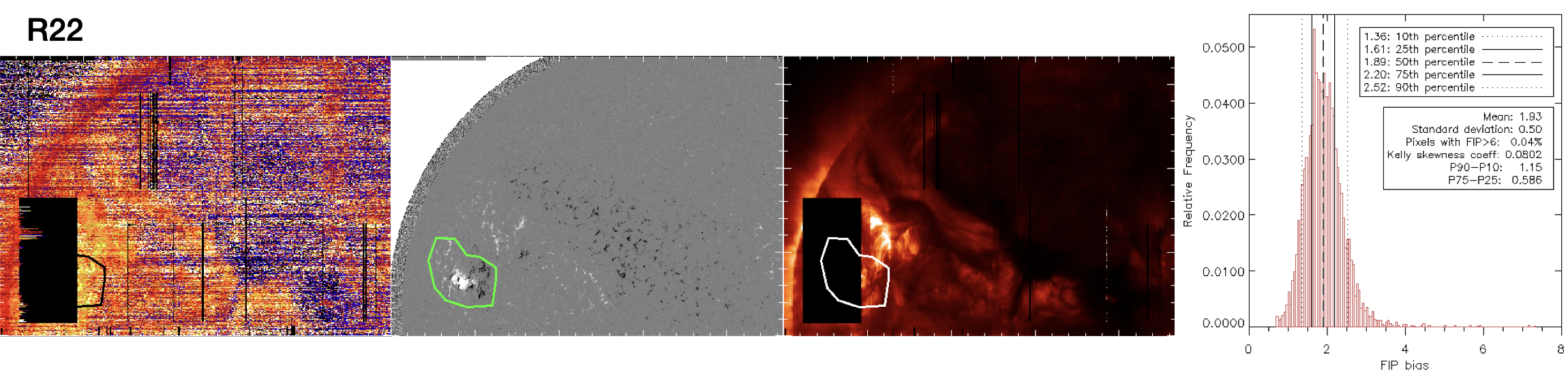}
    \includegraphics[width=1.0\textwidth]{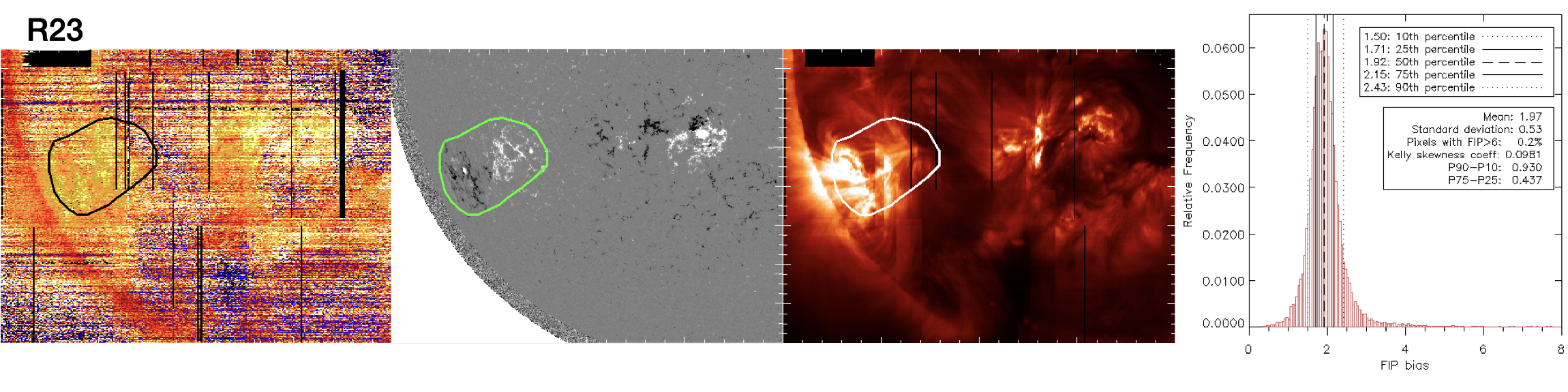}
    \includegraphics[width=1.0\textwidth]{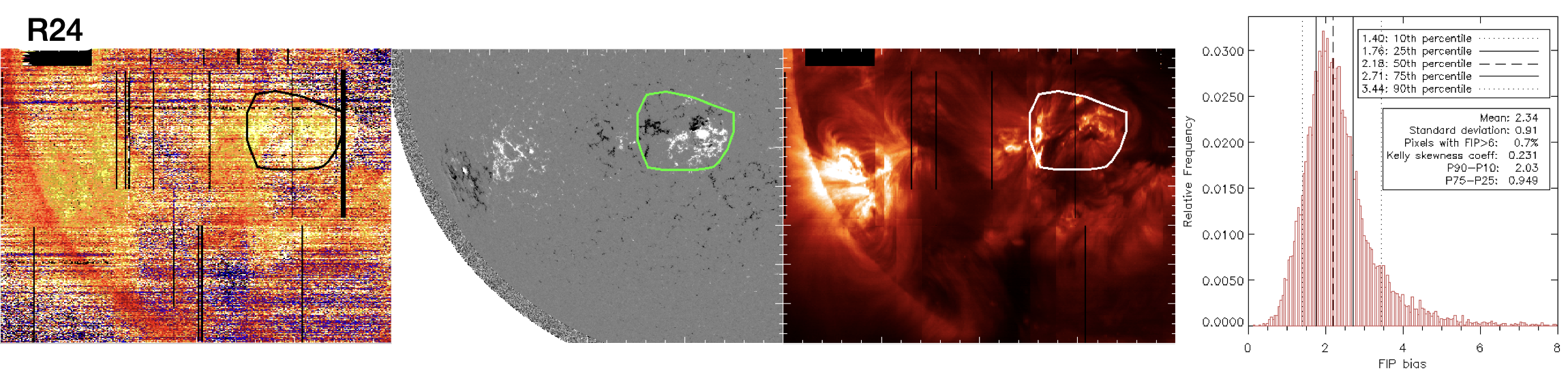}
    \caption{FIP bias distributions in the selected active region contours. From left to right: Hinode EIS FIP bias, HMI line of sight magnetic field strength, Hinode/EIS Fe {\footnotesize XIII} 202.04 \AA{} intensity, and FIP bias distribution within the active region contour (shown in black, green and white respectively).}
    \label{IndividualARContours5}
\end{figure}

\begin{figure}[]
    \centering
    \includegraphics[width=1.0\textwidth]{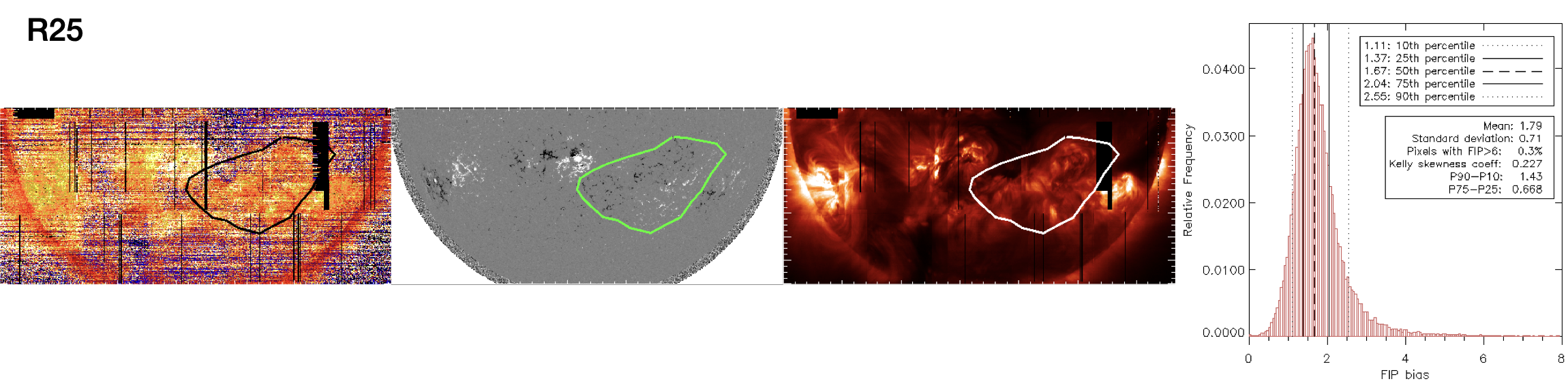}
    \includegraphics[width=1.0\textwidth]{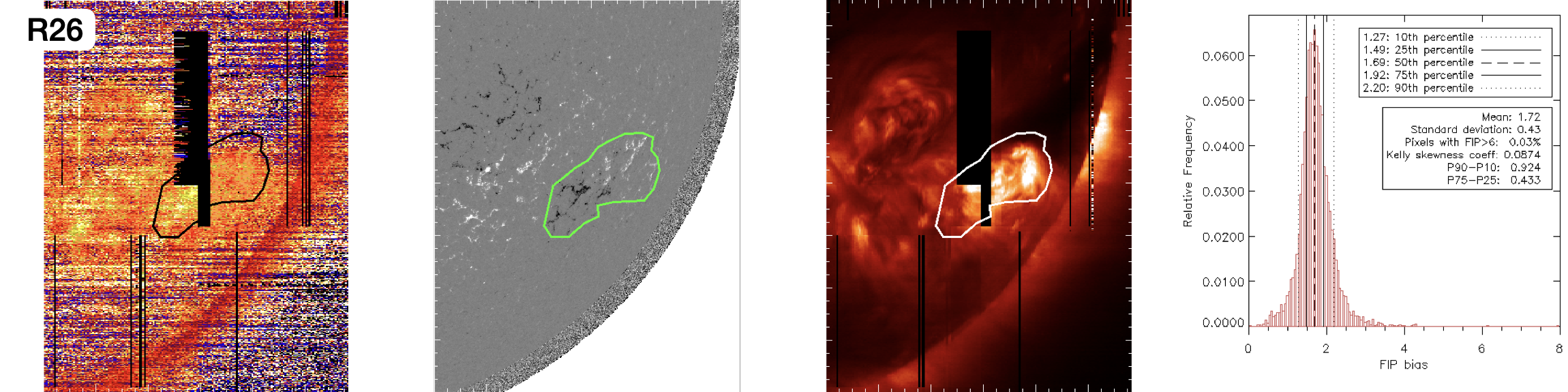}
    \includegraphics[width=1.0\textwidth]{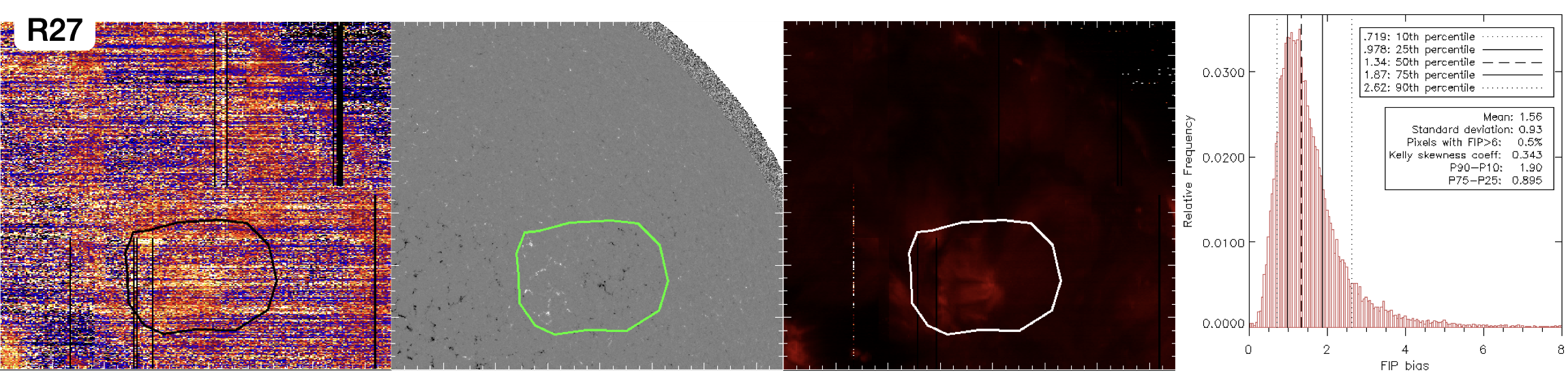}
    \includegraphics[width=1.0\textwidth]{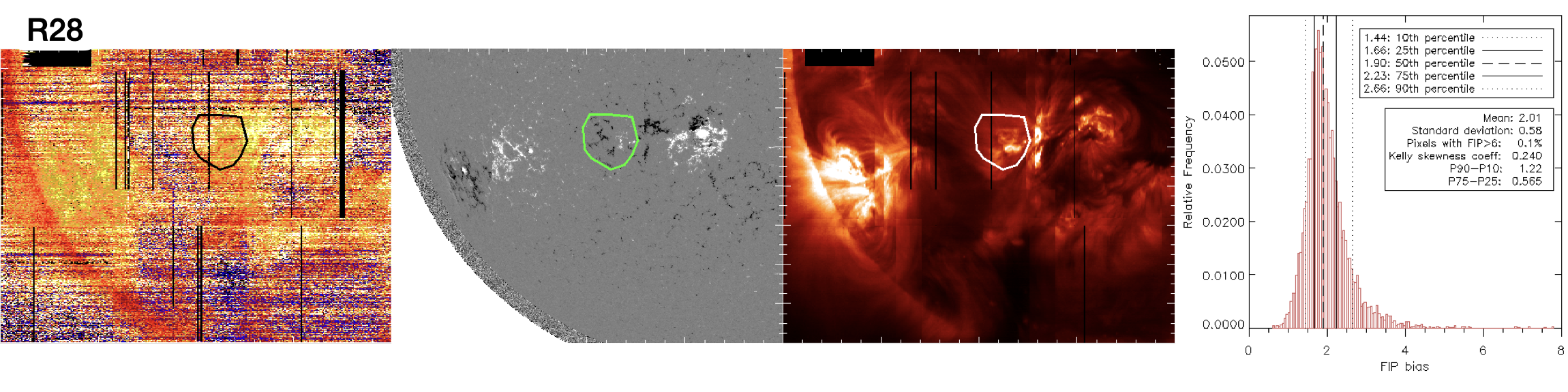}
    \caption{FIP bias distributions in the selected active region contours. From left to right: Hinode EIS FIP bias, HMI line of sight magnetic field strength, Hinode/EIS Fe {\footnotesize XIII} 202.04 \AA{} intensity, and FIP bias distribution within the active region contour (shown in black, green and white respectively).}
    \label{IndividualARContours6}
\end{figure}

\begin{figure}[]
    \centering
    \includegraphics[width=1.0\textwidth]{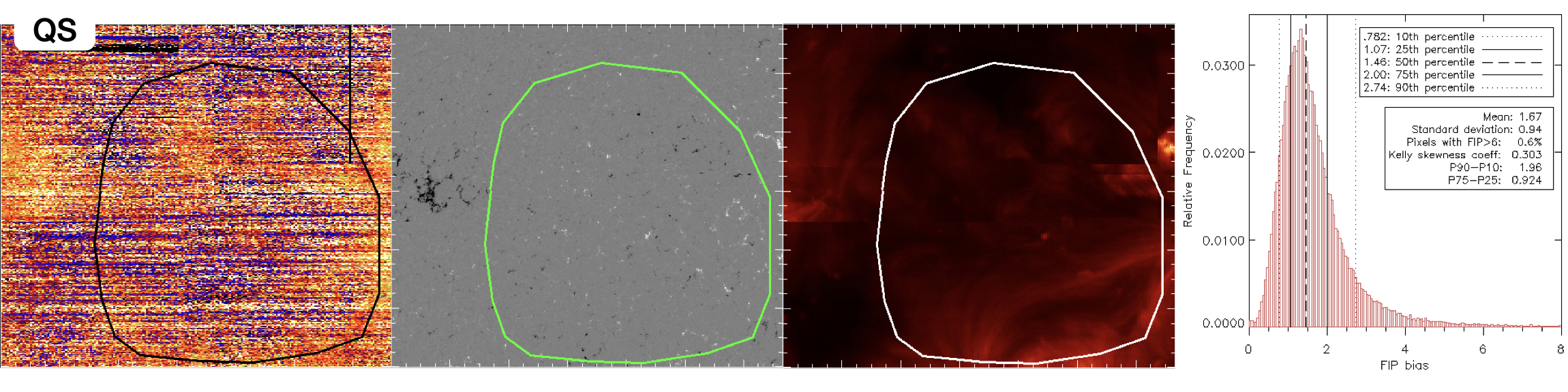}
    \includegraphics[width=1.0\textwidth]{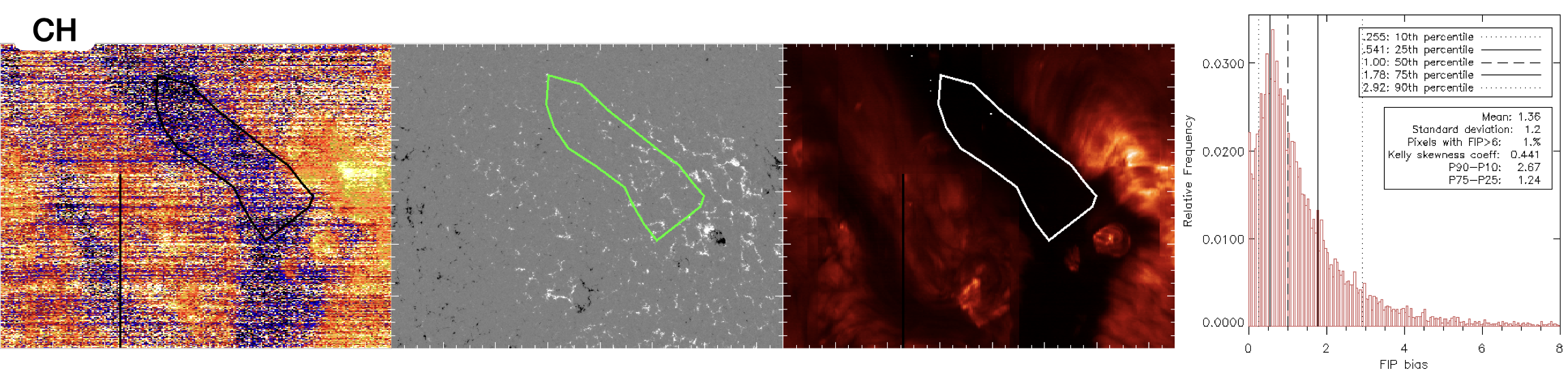}
    \caption{FIP bias distributions in selected representative examples of quiet Sun (QS) and coronal hole (CH). From left to right: Hinode EIS FIP bias, HMI line of sight magnetic field strength, Hinode/EIS Fe {\footnotesize XIII} 202.04 \AA{} intensity, and FIP bias distribution within the selected contour (shown in black, green and white respectively).}
    \label{QSCHcontours}
\end{figure}

\end{document}